\documentclass[a4paper,11pt]{article}
\usepackage{latexsym}
\usepackage{amssymb}
\usepackage{amsfonts}
\usepackage{amsmath}

\renewcommand{\theequation}{\thesection.\arabic{equation}}
\newcommand{\bol}[1]{\boldsymbol {#1}}
\newcommand{\sss}[1]{\scriptscriptstyle {#1}}
\newcommand{\va}{\varphi}
 
\def\d{\partial}

\begin{document}

\author{Nicola Grillo\thanks{\texttt{grillo@physik.unizh.ch}} \\  \\
{\textit{Institut f\"ur Theoretische Physik, Universit\"at Z\"urich}} \\
{\textit{Winterthurerstrasse 190, CH-8057 Z\"urich, Switzerland} }}

\title{Finite One-Loop Calculations in Quantum Gravity: Graviton 
Self-Energy, Perturbative Gauge Invariance and Slavnov--Ward Identities}
\maketitle
\begin{abstract}
\noindent
In this paper we show that the one-loop  graviton self-energy contribution is ultraviolet finite, without 
introducing counterterms, and cutoff-free in the framework of causal perturbation theory. In addition, it 
satisfies the gravitational Slavnov--Ward identities for the two-point connected Green function. The 
condition of perturbative gauge invariance to second order for loop graphs is proved. Corrections to 
the Newtonian potential are also derived.
\vskip 1cm
{\bf PACS numbers:} 0460, 1110
\vskip 7mm
{\bf Keywords:} Quantum Gravity
\vskip 7mm
{\bf Preprint:} ZU-TH 38/1999
\end{abstract}

\newpage
\tableofcontents
\newpage

\section{Introduction}\label{sec:intro}

In the field-theoretical approach to general relativity, Einstein's theory can be reduced to a theory 
of gravitation in flat space-time by an expansion of the gravitational Lagrangian density, the Hilbert--Einstein 
Lagrangian density $\mathcal{L}_{\sss HE}$, as an infinite series in powers of the gravitational coupling constant.

In such an expansion the inherent non-linearity of Einstein equations appears as a non-linear interaction
between gravitons due to their gravitational weight. Thus, Einstein's theory will be considered as  a self-interaction
of ordinary massless rank-2 symmetric tensor gauge fields in flat space-time. The
Lorentz covariant quantization program for the so obtained field theory  was proposed 
in~\cite{gu2},~\cite{fey1},~\cite{ogipu},~\cite{dew3},~\cite{dew4},~\cite{des1} and~\cite{mes2} (and references therein).

For an ample treatment of this  subject, see~\cite{dew5},~\cite{isha2} and~\cite{duf}.

Although covariant Feynman rules in quantum gravity (QG) were soon derived~\cite{fratu},~\cite{fade} which allowed a 
calculable perturbative  expansion of QG~\cite{bro1}, it was soon realized that in the standard 
perturbative framework radiative corrections within the theory were plagued by severe ultraviolet (UV) 
divergences~\cite{hooft},~\cite{velt} (for a review, see~\cite{des7},~\cite{des8},~\cite{des10}).

It turned out that pure (that is without matter fields) QG was one-loop finite due to the Gauss--Bonett  identity in four
dimensions. The obtained one-loop divergences are such that they can be transformed away by a field renormalization.
But two-loop calculations~\cite{goro1},~\cite{goro2} and~\cite{deven}  yield non-renormalizable divergences.

In the meantime, it was realized that the reason for the UV divergences lies basically in the fact that one performs 
mathematically ill-defined operations, when using Feynman rules for closed loop graphs, because one 
multiplies Feynman propagators as if they were ordinary functions.

Therefore a new strategy was developed in order to avoid the appearance of UV divergences once and for all. This was 
done by Epstein and Glaser in the early seventies~\cite{eg}, then further applied to QED by Scharf~\cite{scha} and to 
Yang--Mills theories by D\"utsch \emph{et al.}~\cite{ym1}.

In the resulting scheme, called  `causal perturbation theory', the central object is the $S$-matrix, whose perturbative 
expansion is computed taking causality as cornerstone so that all expressions are finite and well-defined.
UV finiteness is then a consequence of a deeper mathematical understanding of how loop graph contributions have to be 
calculated.

Therefore, power-counting perturbative renormalizability  no longer represents a criterion for distinguishing
viable theories from ill-defined unrealistic theories.

Within the causal perturbation scheme, one-loop contributions to graviton self-energy are calculated in this paper 
and shown to be UV finite without the introduction of a regularization scheme and therefore cutoff-free.
The  obtained graviton self-energy tensor depends logarithmically on a mass scale, which breaks scale invariance,
and satisfies the appropriate gravitational Slavnov--Ward identities~\cite{bro1},~\cite{cap},~\cite{capSW},
only if graviton and ghost loops are added up together.  

An aspect of causal theory applied to QG relies in the fact that one works with free graviton fields in a fixed gauge,
therefore, for  general gauge calculations  we refer to~\cite{cap4},~\cite{cap5}.

Although in this paper we do not present two-loop calculation, the causal method ensures us of their
UV finiteness~\cite{aste}.

For the explicit quantization of the graviton field and the subsequent construction of the physical subspace of the 
graviton Fock space which contains physical graviton states and for the proof of unitarity of the $S$-matrix restricted to
the physical subspace, we refer to~\cite{gri3}, which also provides us with the basic notations and definitions.

QG coupled to photon fields and to scalar matter fields within causal perturbation theory is considered
in~\cite{gri5} and~\cite{gri6}, respectively.

The paper is organized as follow: in the next section, after a brief introduction to causal perturbation theory, the 
transition from general relativity to perturbative quantum gravity in the causal approach is grounded and the condition of 
perturbative gauge invariance is presented and some consequences are drawn. In Sec.~\ref{sec:twopoint} the inductive 
construction of  the graviton 
self-energy is explicitly carried out, the issues of non-normalizability of QG and  distribution splitting 
are touched on. In Sec.~\ref{sec:slav} the Slavnov--Ward identities for the two-point function under investigation are
verified and their relation to perturbative gauge invariance is discussed. In Sec.~\ref{sec:norm} the  freedom in the normalization that
is inherent to the causal inductive construction is settled. Corrections to the Newtonian potential due to  graviton self-energy 
are discussed in Sec.~\ref{sec:potential}, while perturbative gauge invariance to second order in the loop graph sector is 
shown in Sec.~\ref{sec:gauge}.
In the  technical appendices, the formulae needed for the causal construction of the $2$-point distributions
and for the sum of self-energy insertions are derived.

We use the unit convention: $\hbar=c=1$, Greek indices $\alpha,\beta,\ldots$ run from $0$ to $3$, whereas Latin 
indices $i,j,\ldots$ run from $1$ to $3$.

\section{Graviton Coupling and Perturbative Gauge Invariance}\label{sec:grav}
\setcounter{equation}{0}

\subsection{{\boldmath$S$}-Matrix Inductive Construction}\label{sec:smatrix}

The central object of causal perturbation theory~\cite{scha},~\cite{aste},~\cite{gri3} is the scattering matrix $S$.
Being a formal power series in the coupling constant, we consider it  as a sum of smeared operator-valued distributions 
of the following form:
\begin{equation}
S(g)={\mathbf 1}+\sum_{n=1}^{\infty}\frac{1}{n!} \int\!\! d^{4}x_{1}
\ldots d^{4}x_{n}\, T_{n}(x_{1},\ldots, x_{n})\,g(x_{1})\cdot\ldots\cdot g(x_{n})\,  ,
\label{eq:b1}
\end{equation}
where $g$ is a Schwartz test function ($g\in\mathcal{S}(\mathbb{R}^{4})$) which switches the interaction and provides a natural 
infrared cutoff in the long-range part of the interaction.

The $S$-matrix maps the asymptotically incoming free fields on the outgoing ones and it is possible
to express the $T_{n}$'s  by means of free fields. In causal perturbation theory interacting  quantum fields do not
appear.

The $n$-point operator-valued distribution $T_{n}$ is a  well-defined  `renormalized' time-ordered product expressed 
in terms of Wick monomials of free fields.  $T_{n}$ is constructed inductively from the first order $T_{1}(x)$, which 
corresponds to  the interaction Lagrangian in terms of free fields, and from the lower orders $T_{j}$, 
$j=2,\ldots,n-1$ by means of Poincar\'e covariance and causality. 

Causality leads directly to UV finite and cutoff-free $T_{n}$-distributions in every order without introducing any counterterm. 



\subsection{First Order Graviton Interaction}\label{sec:first}


Following the usual approach~\cite{bro1},~\cite{cap}, we start from the Hilbert--Einstein Lagrangian (without 
cosmological constant)
\begin{equation}
\mathcal{L}_{\sss HE}={\frac{- 2}{\kappa ^ 2}}\sqrt{-g} R\, ,
\label{eq:b2}
\end{equation}
where $R$ is the Ricci scalar and $\kappa^2=32\,\pi\, G$ with $G=$ Newton's constant. We use the same notations
as in~\cite{gri3} and~\cite{scho1}. Expanding  the Goldberg variable $\tilde{g}^{\mu\nu}:=\sqrt{-g}\, g^{\mu\nu}$ in an 
asymptotically flat geometry
\begin{equation}
\tilde{g}^{\mu\nu}(x)=\eta^{\mu\nu}+\kappa\, h^{\mu\nu}(x)\, ,
\label{eq:b3}
\end{equation}
where $\eta^{\mu\nu}=\mathrm{diag}(1,-1,-1,-1)$ is the flat space-time metric tensor, we find the non-terminating 
expansion  of $\mathcal{L}_{\sss HE}$
\begin{equation}
\mathcal{L}_{\sss HE}=\sum_{j=0}^{\infty}\, \kappa^{j}\, \mathcal{L}_{\sss HE}^{\sss{(j)}}\, ,
\label{eq:b4}
\end{equation}
where $\mathcal{L}_{\sss HE}^{\sss{(j)}}$ represents an interaction involving $j+2$ gravitons.  
Eq.~(\ref{eq:b3}) defines the dynamical graviton field $h^{\mu\nu}(x)$ propagating in the flat space-time geometry. 

The lowest order $\mathcal{L}_{\sss HE}^{\sss{(0)}}$ is quadratic in $h^{\mu\nu}(x)$  and in the Hilbert gauge
$h^{\mu\nu}(x)_{,\nu}=0$  the graviton field $h^{\mu\nu}(x)$ obeys the wave equation
\begin{equation}
\Box \, h^{\mu\nu}(x)=0\, .
\label{eq:b5}
\end{equation}

Since the perturbative expansion for the $S$-matrix~(\ref{eq:b1}) is in powers of the coupling constant $\kappa$,
we consider the normally ordered product of the first order term in~(\ref{eq:b4})
\begin{multline}
T_{1}^{h}(x) =i\, \kappa :\!{\mathcal{L}}_{\sss HE}^{ \sss{(1)}}(x)\!: 
             =i\,  \frac{\kappa}{2}  \big\{
                  + :\!h^{\rho\sigma}h^{\alpha\beta}_{\: ,\rho}h^{\alpha\beta}_{\:,\sigma}\!:
       -\frac{1}{2}:\!h^{\rho\sigma}h_{,\rho}h_{,\sigma}\!:+  \\
       +2:\!h^{\rho\sigma}h^{\sigma\beta}_{\:,\alpha}h^{\rho\alpha}_{\:,\beta}\!:
                  +:\!h^{\rho\sigma}h_{\:,\alpha}h^{\rho\sigma}_{\:,\alpha}\!:
       -2:\!h^{\rho\sigma}h^{\alpha\rho}_{\:,\beta}h^{\sigma\alpha}_{\:,\beta}\!: \big\}\,,
\label{eq:b6}
\end{multline}
as the  cubic interaction between  gravitons or first order graviton interaction. 
For brevity, we omit the space-time dependence of the fields, if the meaning is clear.

For convenience of notation, the trace of the graviton field is written as $h=h^{\gamma}_{\  \gamma}$ and all Lorentz 
indices of the graviton fields are written as superscripts whereas the derivatives acting on the fields are written as 
subscripts. All indices occurring twice are contracted by the Minkowski metric $\eta^{\mu\nu}$.


The `non-renormalizability problem'  of quantum gravity arises because  of the  presence of two derivatives on the 
graviton fields in~(\ref{eq:b6}) whose origin lies in the dimensionality of the coupling constant: 
$[\kappa ]=\mathrm{mass}^{-1}$.

\subsection{Quantization of the Graviton Field, Perturbative Gauge Invariance and Ghost Coupling}\label{sec:quant}

We consider the graviton field $h^{\mu\nu}(x)$ as a free quantum tensor field which satisfies the wave 
equation~(\ref{eq:b5}) and quantize it by imposing the Lorentz covariant commutation rule
\begin{equation}
\big[ h^{\alpha\beta}(x),h^{\mu\nu}(y) \big]=-i\, b^{\alpha\beta\mu\nu}\, D_{0}(x-y)\, ,
\label{eq:b7}
\end{equation}
with
\begin{equation}
b^{\alpha\beta\mu\nu}:=\frac{1}{2}\left(  \eta^{\alpha\mu}\eta^{\beta\nu}+ 
  \eta^{\alpha\nu}\eta^{\beta\mu}-\eta^{\alpha\beta}\eta^{\mu\nu}\right)\, ;
\label{eq:b8}
\end{equation}
and $D_{0}(x)$ is the mass-zero Jordan--Pauli causal distribution:
\begin{equation}
\begin{split}
D_{0}(x)&=D^{\sss (+)}_{0}(x)+D^{\sss (-)}_{0}(x)=\frac{1}{2\pi}\delta(x^2)\, 
\mathrm{sgn}(x^{0}) \\
&=\frac{i}{(2\pi)^3}\int \!\!d^{4}p\, \delta (p^2)\,
     \mathrm{sgn}(p^{0})\, e^{-i\,p\cdot x} \, .
\label{eq:b9}
\end{split}
\end{equation}

The gauge content of quantum gravity is formulated by means of the Lorentz invariant and time-independent gauge charge $Q$:
\begin{equation}
Q:=\int\limits_{x^0 =const}\!\! d^{3}x\,h^{\alpha\beta}(x)_{,\beta} {\stackrel{\longleftrightarrow}{ \partial_{x}^{0} }
 u_{\alpha}(x)} \, ,
\label{eq:b10}
\end{equation}
where $u^{\mu}(x)$ is a C-number vector field satisfying $\Box u^{\mu}(x)=0$. 
The gauge charge generates the 
infinitesimal operator gauge transformation of the graviton field $h^{\mu\nu}(x)$:
\begin{equation}
d_Q h^{\mu\nu}(x) : =\big[ Q, h^{\mu\nu}(x)\big]=-i\, b^{\mu\nu\rho\sigma} u^{\rho}(x)_{,\sigma}\, .
\label{eq:b11.1}
\end{equation}
The ten components of the symmetric rank-2 tensor  $h^{\mu\nu}$ contain more than the true physical degrees of freedom 
of a massless spin-2 particle, this  additional freedom could be suppressed by a gauge  condition $h^{\mu\nu}_{\: ,\nu}=0$ 
and a trace condition $h=0$. As in gauge theories these conditions are disregarded at the beginning and considered later 
as conditions on the physical states. In~\cite{gri3} the explicit construction of the Fock space for the physical 
graviton states is carried out and there it is shown that the physical subspace can be defined as 
$\mathcal{F}_{phys}=\ker\big(\{ Q,Q^{\dagger}\}\big)$.

Gauge invariance of the $S$-matrix means
\begin{equation}
\lim_{g\to 1}\big[ Q, S(g) \big]=0\,.
\label{eq:b12}
\end{equation}
This condition is fulfilled, if  the perturbative gauge invariance condition for the $n$-point distribution $T_{n}$
\begin{equation}
d_Q T_{n}(x_1 ,\ldots,x_n )=\text{sum of divergences}\, ,
\label{eq:b13}
\end{equation}
holds, because divergences do not contribute in the adiabatic limit $g \to 1$ due to partial integration and Gauss' theorem.

For $n=1$ the above requirement is not at all trivial, because   $d_Q T_{1}^{h}(x)\neq divergence$.
This requires the introduction of an interaction between graviton, ghost and anti-ghost, \emph{i.e.}
the  first order ghost coupling~\cite{kugo},~\cite{nishi}
\begin{multline}
T_{1}^{u}=i\, \kappa\, \big( +:\!\tilde{u}^{\nu}_{,\mu} h^{\mu\nu}_{,\rho} u^{\rho}\!:
                             -:\!\tilde{u}^{\nu}_{,\mu} h^{\nu\rho}        u^{\mu}_{,\rho}\!: 
                             -:\!\tilde{u}^{\nu}_{,\mu} h^{\mu\rho}        u^{\nu}_{,\rho}\!:
                             +:\!\tilde{u}^{\nu}_{,\mu} h^{\mu\nu}         u^{\rho}_{,\rho}\!: \big)\, .
\label{eq:b14}
\end{multline}
The ghost fields must be quantized as free fermionic vector fields
\begin{equation}
\Box u^{\nu}(x)=0\,,\quad \Box \tilde{u}^{\nu}(x)=0\,,\quad 
\big\{u^{\mu}(x),\tilde{u}^{\nu}(y)\big\}=i\, \eta^{\mu\nu}\, D_{0}(x-y) \, ,
\label{eq:b15}
\end{equation}
whereas all other anti-commutators vanish. The ghost and anti-ghost fields undergo the infinitesimal operator gauge 
variations
\begin{equation}
d_Q u^{\alpha}(x) := \big\{ Q, u^{\alpha}(x)\big\}=0\,,\quad
d_Q \tilde{u}^{\alpha}(x) := \big\{ Q, \tilde{u}^{\alpha}(x)\big\}=i\, h^{\alpha\beta}(x)_{,\beta}
\label{eq:b11.2}
\end{equation}
under the action of $Q$, so that the sum of~(\ref{eq:b6}) and~(\ref{eq:b14}) preserves perturbative gauge 
invariance~(\ref{eq:b13}) to first order:
\begin{equation}
d_{Q} T_{1}^{h+u}(x)=d_{Q}\big(T_{1}^{h}(x)+T_{1}^{u}(x)\big)=:\partial_{\nu}^{x}T_{1/1}^{ \nu}(x)\,.
\label{eq:b16}
\end{equation}
One possible  form of $T_{1/1}^{\nu}(x)$, the so-called $Q$-vertex, was derived in~\cite{scho1}. 

The definition of the $Q$-vertex from Eq.~(\ref{eq:b16}) allows us to give  a precise prescription on how the right side of
Eq.~(\ref{eq:b13}) has to be inductively constructed.  We define the concept of `perturbative quantum operator gauge 
invariance' by the equation
\begin{equation}
d_{Q}T_{n}(x_{1},\ldots ,x_{n})=\sum_{l=1}^{n}\frac{\partial}{\partial x^{\nu}_{l}}\, 
T_{n/l}^{\nu}(x_{1},\ldots , x_{l},\ldots ,x_{n}) \, ,
\label{eq:b17}
\end{equation}
where  $T_{n/l}^{\nu}$ is the `renormalized' time-ordered product, obtained according to the inductive causal scheme, with a
$Q$-vertex at $x_{l}$, while all other $n-1$ vertices are ordinary $T_{1}$-vertices.

\subsection{Consequences of Perturbative Gauge Invariance to Second Order for Two-Point Distributions}

We derive now some consequences from the condition of perturbative gauge invariance to second order for loop graphs.
From the structure of $T_{1}=T_{1}^{h}+T_{1}^{u}$, it follows straightforwardly that by performing two field 
contractions~(\ref{eq:b18}) the resulting $2$-point distribution $T_{2}(x,y)$ will be of the form
\begin{equation}
\begin{split}
T_{2}(x,y)=
&+:\!h^{\alpha\beta}(x)h^{\mu\nu}(y)\!:\,i\,t_{\sss hh}(z)^{\alpha\beta\mu\nu}
+:\!h^{\alpha\beta}(x)_{,\epsilon}h^{\mu\nu}(y)\!:\,i\,t_{\sss \d hh}(z)_{\epsilon|}^{\alpha\beta\mu\nu}+\\
&+:\!h^{\alpha\beta}(x)h^{\mu\nu}(y)_{,\rho}\!:\,i\,t_{\sss h\d h}(z)_{\ |\rho}^{\alpha\beta\mu\nu}
+:\!h^{\alpha\beta}(x)_{,\epsilon}h^{\mu\nu}(y)_{,\rho}\!:\,i\,
             t_{\sss \d h\d h}(z)_{\epsilon|\rho}^{\alpha\beta\mu\nu}\\
&+:\!u^{\gamma}(x)\tilde{u}^{\mu}(y)_{,\rho}\!:\,i\,t_{\sss u\d\tilde{u}}(z)^{\gamma|\mu}_{\ |\rho}
+:\!\tilde{u}^{\alpha}(x)_{,\beta}{u}^{\delta}(y)\!:\,i\,
          t_{\sss \d \tilde{u}{u}}(z)^{\alpha|\delta}_{\beta|}+\\
&+:\!u^{\gamma}(x)_{,\delta}\tilde{u}^{\mu}(y)_{,\rho}\!:\,i\,
                   t_{\sss \d u\d\tilde{u}}(z)^{\gamma|\mu}_{\delta|\rho}
+:\!\tilde{u}^{\mu}(x)_{,\nu}{u}^{\alpha}(y)_{,\beta}\!:\,i\,
               t_{\sss \d\tilde{u}\d{u}}(z)^{\mu|\alpha}_{\nu|\beta}\,,
\raisetag{13mm}\label{eq:b17.1}
\end{split}
\end{equation}
where $z:=x-y$. The subscript on the numerical $t$-distribution denotes the structure of the external fields attached to them. 
This $T_{2}(x,y)$ describes the graviton self-energy and the ghost self-energy. The corresponding tensors will be given in 
Sec.~\ref{sec:selfenergy} and in Sec.~\ref{sec:ghose}, respectively.

Perturbative gauge invariance to second order
\begin{equation}
d_{Q}T_{2}(x,y)=\d_{\nu}^{x} T_{2/1}^{\nu}(x,y) +\d_{\nu}^{y} T_{2/2}^{\nu}(x,y)
\label{eq:b17.2}
\end{equation}
enables us to derive a set of identities for these distributions by comparing the distributions attached to  same external 
operators on both sides of~(\ref{eq:b17.2}).

The left side of~(\ref{eq:b17.2}) is obtained by calculating the infinitesimal gauge variations of the external 
fields\footnote{$d_{Q}(:\!u\tilde{u}\!:)=-:\!u\{Q,\tilde{u}\}\!:=-:\!uh\!:$ and 
$d_{Q}(:\!\tilde{u}u\!:)=:\!\{Q,\tilde{u}\}u\!:=:\!hu\!:$}
by means of Eqs.~(\ref{eq:b11.1}) and~(\ref{eq:b11.2}) and we  isolate the  terms with external operators of the type 
$:\!u(x)h(y)\!:$ so that
\begin{equation}
\begin{split}
d_{Q}&T_{2}(x,y)=
+b^{\alpha\beta\gamma\delta}\,\Big[:\!u^{\gamma}(x)_{,\delta}h^{\mu\nu}(y)\!:
                         t_{\sss hh}(z)^{\alpha\beta\mu\nu}
+:\!u^{\gamma}(x)_{,\delta,\epsilon}h^{\mu\nu}(y)\!:
                              t_{\sss \d h h}(z)_{\epsilon|}^{\alpha\beta\mu\nu}\\
&+:\!u^{\gamma}(x)_{,\delta}h^{\mu\nu}(y)_{,\rho}\!:\,
                 t_{\sss h\d h}(z)_{\ |\rho}^{\alpha\beta\mu\nu}
+:\!u^{\gamma}(x)_{,\delta,\epsilon}h^{\mu\nu}(y)_{,\rho}\!:\,
                  t_{\sss \d h\d h}(z)_{\epsilon|\rho}^{\alpha\beta\mu\nu}\Big]+\\
&+:\!u^{\gamma}(x)h^{\mu\nu}(y)_{,\rho,\tau}\!:\,t_{\sss u\d\tilde{u}}(z)^{\gamma|\mu}_{\ |\rho}\,
                         \eta^{\nu\tau}
+:\!u^{\gamma}(x)_{,\delta}h^{\mu\nu}(y)_{,\rho,\tau}\!:\,
      t_{\sss \d u\d\tilde{u}}(z)^{\gamma|\mu}_{\delta|\rho}\,\eta^{\nu\tau}+\ldots\,.
\raisetag{12mm}\label{eq:b17.3}
\end{split}
\end{equation}
On the other side, $\d_{\nu}^{x} T_{2/1}^{\nu} +\d_{\nu}^{y} T_{2/2}^{\nu}$ contains also operators of this type. Using a simplified
notation which keeps track of the field type, of the derivatives and of the position of the $\nu$-index which forms the divergence 
in~(\ref{eq:b17.2}), then the $Q$-vertex of~(\ref{eq:b16}) reads (see~\cite{scho1} for the detailed form):
\begin{equation}
\begin{split}
T_{1/1}^{\nu}(x):=&+:\!\d u h \d^{x}_{\nu}h \!: +:\!u \d h \d^{x}_{\nu}h \!:  +:\!u^{\nu}\d h\d h \!:   
                  +:\!\d u h \d h^{\nu} \!:+\\&+:\!\d u^{\nu} h \d h \!: +:\!u \d h \d h^{\nu} \!: 
          +:\!\d u\d h h^{\nu} \!: +:\!\d^{x}_{\nu}\tilde{u} u \d u \!:+\\&+:\!\tilde{u} \d u \d u^{\nu} \!:
            +:\!\tilde{u} u \d\d u^{\nu} \!:   +:\!\tilde{u}^{\nu} u \d\d u \!: \,,
\label{eq:b17.4}
\end{split}
\end{equation}
while the `normal' vertex reads
\begin{equation}
T_{1}:=:\!h\d h\d h\!:+:\!\d \tilde{u}\d h u\!:+:\!\d \tilde{u} h \d u\!:\,.
\label{eq:b17.5}
\end{equation}
Then, performing two contractions between $T_{1/1}^{\nu}(x)$ and $T_{1}(y)$, the contributions  in $T_{2/1}^{\nu}$ which have
external operators of the type $:\!u(x)h(y)\!:$ are
\begin{equation}
\begin{split}
T_{2/1}^{\nu}(x,y)=
&+:\!u^{\gamma}(x)_{,\delta}h^{\mu\nu'}(y)\!:\,t^{\nu}_{\sss \d uh}(z)^{\gamma|\mu\nu'}_{\delta|}
+:\!u^{\gamma}(x)_{,\delta}h^{\mu\nu'}(y)_{,\rho}\!:\,
           t^{\nu}_{\sss \d u\d h}(z)^{\gamma|\mu\nu'}_{\delta|\rho}+\\
&+:\!u^{\gamma}(x)h^{\mu\nu'}(y)\!:\,
           t^{\nu}_{\sss u h}(z)^{\gamma|\mu\nu'}
+:\!u^{\gamma}(x)h^{\mu\nu'}(y)_{,\rho}\!:\,
           t^{\nu}_{\sss u\d h}(z)^{\gamma|\mu\nu'}_{\ |\rho}+\\
&+:\!u^{\nu}(x)h^{\mu\nu'}(y)\!:\,t_{\sss uh}(z)^{|\mu\nu'}
 +:\!u^{\nu}(x)h^{\mu\nu'}(y)_{,\rho}\!:\,t_{\sss u\d h}(z)^{|\mu\nu'}_{\ |\rho}+\\
&+:\!u^{\nu}(x)_{,\delta}h^{\mu\nu'}(y)\!:\,t_{\sss \d u h}(z)^{\ |\mu\nu'}_{\delta|}
+:\!u^{\nu}(x)_{,\delta}h^{\mu\nu'}(y)_{,\rho}\!:\,t_{\sss \d u\d h}(z)^{\ |\mu\nu'}_{\delta|\rho}+\\
+:\!u^{\nu}(x)_{,\delta,\epsilon}&h^{\mu\nu'}(y)_{,\rho}\!:\,
                           t_{\sss \d\d u\d h}(z)^{\ \: |\mu\nu'}_{\delta\epsilon|\rho}
+:\!u^{\nu}(x)_{,\delta,\epsilon}h^{\mu\nu'}(y)\!:\,
                           t_{\sss \d\d u  h}(z)^{\ \: |\mu\nu'}_{\delta\epsilon|}+\\
+:\!u^{\gamma}(x)_{,\delta,\epsilon}&h^{\mu\nu'}(y)_{,\rho}\!:\,
           t^{\nu}_{\sss \d\d u\d h}(z)^{\ \gamma|\mu\nu'}_{\delta\epsilon|\rho}
+:\!u^{\gamma}(x)_{,\delta,\epsilon}h^{\mu\nu'}(y)\!:\,
           t^{\nu}_{\sss \d\d u  h}(z)^{\ \gamma|\mu\nu'}_{\delta\epsilon|}\,.
\raisetag{11mm}\label{eq:b17.6}
\end{split}
\end{equation}
One should not forget that there exist also terms with external operators of the type $:\!u(x)h(y)\!:$ coming from $T_{2/2}^{\nu}(x,y)$,
which is inductively constructed with $T_{1}(x)$ and $T_{1/1}^{\nu}(y)$. They read
\begin{equation}
\begin{split}
T_{2/2}^{\nu}(x,y)=
&+:\!u^{\gamma}(x)h^{\mu\nu'}(y)\!:\,
           l^{\nu}_{\sss u h}(z)^{\gamma|\mu\nu'}
+:\!u^{\gamma}(x)h^{\mu\nu'}(y)_{,\nu}\!:\,
           l_{\sss u\d h}(z)^{\gamma|\mu\nu'}+\\
&+:\!u^{\gamma}(x)h^{\mu\nu'}(y)_{,\rho}\!:\,
           l^{\nu}_{\sss u\d h}(z)^{\gamma|\mu\nu'}_{\ |\rho}
+:\!u^{\gamma}(x)h^{\mu\nu}(y)_{,\rho}\!:\,
           \hat{l}_{\sss u\d h}(z)^{\gamma|\mu}_{\ |\rho}+\\
&+:\!u^{\gamma}(x)h^{\mu\nu}(y)\!:\,
           {l}_{\sss u h}(z)^{\gamma|\mu}
+:\!u^{\gamma}(x)_{,\delta}h^{\mu\nu'}(y)\!:\,l^{\nu}_{\sss \d uh}(z)^{\gamma|\mu\nu'}_{\delta|}+\\
&+:\!u^{\gamma}(x)_{,\delta}h^{\mu\nu'}(y)_{,\nu}\!:\,
           l_{\sss \d u\d h}(z)^{\gamma|\mu\nu'}_{\delta|}
+:\!u^{\gamma}(x)_{,\delta}h^{\mu\nu'}(y)_{,\rho}\!:\,
           l^{\nu}_{\sss \d u\d h}(z)^{\gamma|\mu\nu'}_{\delta|\rho}\\
&+:\!u^{\gamma}(x)_{,\delta}h^{\mu\nu}(y)_{,\rho}\!:\,
           \tilde{l}_{\sss \d u\d h}(z)^{\gamma|\mu}_{\delta|\rho}
+:\!u^{\gamma}(x)_{,\delta}h^{\mu\nu}(y)\!:\,
           {l}_{\sss \d u h}(z)^{\gamma|\mu}_{\delta|}\,.
\label{eq:b17.7}
\end{split}
\end{equation}
Here, the numerical distributions are denoted by $l$. According to Eq.~(\ref{eq:b17.2}), we have to apply $\d_{\nu}^{x}$ to 
$T_{2/1}^{\nu}$,~(\ref{eq:b17.6}), and $\d_{\nu}^{y}$ to $T_{2/2}^{\nu}$,~(\ref{eq:b17.7}). After that, we gather the various terms 
according to their Lorentz structures given by the position of the indices and the number of derivatives acting on the external 
fields. We compare then the C-number distributions attached to the external operators:
\begin{gather}
:\!u^{\gamma}(x)_{,\delta}h^{\mu\nu}(y)\!:\,,\quad
:\!u^{\gamma}(x)_{,\delta,\epsilon}h^{\mu\nu}(y)\!:\,,\quad
:\!u^{\gamma}(x)_{,\delta}h^{\mu\nu}(y)_{,\rho}\!:\,,\nonumber\\
:\!u^{\gamma}(x)_{,\delta,\epsilon}h^{\mu\nu}(y)_{,\rho}\!:\,,\quad
:\!u^{\gamma}(x)h^{\mu\nu}(y)_{,\rho,\tau}\!:\,,\quad
:\!u^{\gamma}(x)_{,\delta}h^{\mu\nu}(y)_{,\rho,\tau}\!:\,,\quad
:\!u^{\gamma}(x)h^{\mu\nu}(y)\!:,\nonumber\\
:\!u^{\gamma}(x)h^{\mu\nu}(y)_{,\rho}\!:\,,\quad
:\!u^{\gamma}(x)_{,\delta,\epsilon,\psi}h^{\mu\nu}(y)_{,\rho}\!:\,,\quad
:\!u^{\gamma}(x)_{,\delta,\epsilon,\psi}h^{\mu\nu}(y)\!:\,,
\label{eq:b17.8}
\end{gather}
between $d_{Q}T_{2}$ and $\d_{\nu}^{x}T_{2/1}^{\nu}+\d_{\nu}^{y}T_{2/2}^{\nu}$ so that we  obtain the identities
\begin{equation}
\begin{split}
&\begin{split}(i1),\quad b^{\alpha\beta\gamma\delta}\,t_{\sss hh}(z)^{\alpha\beta\mu\nu}=
   \bigg\{ &+ \d_{\eta}^{x} t^{\eta}_{\sss \d uh}(z)^{\gamma|\mu\nu}_{\delta|}+  
        t^{\delta}_{\sss u h}(z)^{\gamma|\mu\nu}+\eta^{\gamma\delta}\,t_{\sss uh}(x)^{|\mu\nu} + \\
          &+ \d_{\gamma}^{x}\,t_{\sss \d u h}(z)^{\ |\mu\nu}_{\delta|} +
              \d_{\pi}^{y}\,l^{\pi}_{\sss \d uh}(z)^{\gamma|\mu\nu}_{\delta|}+
       \d_{\nu}^{y}\,{l}_{\sss \d u h}(z)^{\gamma|\mu}_{\delta|}    \bigg\}\,,\end{split}\\
&\begin{split}(i2),\quad b^{\alpha\beta\gamma\delta}\,t_{\sss \d h h}(z)_{\epsilon|}^{\alpha\beta\mu\nu}=
       \bigg\{&+t^{\epsilon}_{\sss \d uh}(z)^{\gamma|\mu\nu}_{\delta|}+
                  \eta^{\gamma\epsilon}\, t_{\sss \d u h}(z)^{\ |\mu\nu}_{\delta|} +
               \d_{\psi}^{x}\,t^{\psi}_{\sss \d\d u  h}(z)^{\gamma|\mu\nu}_{\delta\epsilon|}+\\
         &+   \d_{\gamma}^{x}\, t_{\sss \d\d u  h}(z)^{\ \: |\mu\nu}_{\delta\epsilon|} \bigg\}\,,\end{split}\\
&\begin{split}(i3),\quad b^{\alpha\beta\gamma\delta}\,t_{\sss h\d h}(z)_{\ |\rho}^{\alpha\beta\mu\nu}=
        \bigg\{&+\d_{\eta}^{x}\,t^{\eta}_{\sss \d u\d h}(z)^{\gamma|\mu\nu}_{\delta|\rho} +
          t^{\delta}_{\sss u\d h}(z)^{\gamma|\mu\nu}_{\ |\rho}+
           \d_{\gamma}^{x}\,t_{\sss \d u\d h}(z)^{\ |\mu\nu}_{\delta|\rho} +\\
          &+\eta^{\gamma\delta}\, t_{\sss u\d h}(z)^{|\mu\nu}_{\ |\rho}  +
           l^{\rho}_{\sss \d uh}(z)^{\gamma|\mu\nu}_{\delta|}+
           \d_{\rho}^{y}\,l_{\sss \d u\d h}(z)^{\gamma|\mu\nu}_{\delta|}+\\
          &+\d_{\pi}^{y}\,l^{\pi}_{\sss \d u \d h}(z)^{\gamma|\mu\nu}_{\delta|\rho}+
           \d_{\nu}^{y}\,\tilde{l}_{\sss \d u \d h}(z)^{\gamma|\mu}_{\delta|\rho}+
              \eta^{\nu\rho}\, l_{\sss \d u h}(z)^{\gamma|\mu}_{\delta|}\bigg\}\,,\end{split}\\
&\begin{split}(i4),\quad b^{\alpha\beta\gamma\delta}\,t_{\sss \d h\d h}(z)_{\epsilon|\rho}^{\alpha\beta\mu\nu}=
           \bigg\{&+t^{\epsilon}_{\sss \d u\d h}(z)^{\gamma|\mu\nu}_{\delta|\rho}+
                  \eta^{\gamma\epsilon}\, t_{\sss \d u\d h}(z)^{\ |\mu\nu}_{\delta|\rho} +
               \d_{\psi}^{x}\,t^{\psi}_{\sss \d\d u \d h}(z)^{\gamma|\mu\nu}_{\delta\epsilon|\rho}\\
       &+\d_{\gamma}^{x}\, t_{\sss \d\d u \d h}(z)^{\ |\mu\nu}_{\delta\epsilon|\rho}\bigg\}\,,\end{split}\\
&\begin{split}(i5),\quad t_{\sss u\d \tilde{u}}(z)^{\gamma|\mu}_{\ |\rho}\,\eta^{\nu\tau}=
        \bigg\{&+\eta^{\rho\tau}\,l_{\sss u\d h}(z)^{\gamma|\mu\nu}+
           l^{\tau}_{\sss u\d h}(z)^{\gamma|\mu\nu}_{\ |\rho}+
           \eta^{\nu\tau}\,\hat{l}_{\sss u\d h}(z)^{\gamma|\mu}_{\ |\rho}\bigg\}\,,\end{split}\\
&\begin{split}(i6),\quad t_{\sss \d u\d\tilde{u}}(z)^{\gamma|\mu}_{\delta|\rho}\,\eta^{\nu\tau}=
      \bigg\{&+\eta^{\rho\tau}\,l_{\sss \d u\d h}(z)^{\gamma|\mu\nu}_{\delta|}+
           l^{\tau}_{\sss \d u\d h}(z)^{\gamma|\mu\nu}_{\delta|\rho}+
           \eta^{\nu\tau}\,\tilde{l}_{\sss \d u\d h}(z)^{\gamma|\mu}_{\delta|\rho} \bigg\}\,,\end{split}\\
&\begin{split}(i7),\quad 0=\bigg\{&+\d_{\eta}^{x}t^{\eta}_{\sss uh}(z)^{\gamma|\mu\nu}+
                      \d_{\gamma}^{x}t_{\sss uh}(z)^{|\mu\nu}
            +\d_{\pi}^{y}l^{\pi}_{\sss uh}(z)^{\gamma|\mu\nu} +
             \d_{\nu}^{y}l_{\sss uh}(z)^{\gamma|\mu}\bigg\}\,,\end{split}\\
&\begin{split}(i8),\quad 0=\bigg\{&+\d_{\eta}^{x}t^{\eta}_{\sss u\d h}(z)^{\gamma|\mu\nu}_{\ |\rho}+
     \d_{\gamma}^{x}t_{\sss u\d h}(z)^{|\mu\nu}_{|\rho}+
     l^{\rho}_{\sss uh}(z)^{\gamma|\mu\nu}+\d_{\rho}^{y}l_{\sss u\d h}(z)^{\gamma|\mu\nu}\\
           &+\d_{\pi}^{y}l^{\pi}_{\sss u\d h}(z)^{\gamma|\mu\nu}_{\ |\rho}+
      \d_{\nu}^{y}\hat{l}_{\sss u\d h}(z)^{\gamma|\mu}_{\ |\rho}+
      \eta^{\nu\rho}l_{\sss uh}(z)^{\gamma|\mu}\bigg\}\,,\end{split}\\
&\begin{split}(i9),\quad 0=\bigg\{&+\eta^{\psi\gamma}\,t_{\sss \d\d u\d  h}(z)^{\ \: |\mu\nu}_{\delta\epsilon|\rho}+
           t^{\psi}_{\sss \d\d u\d h}(z)^{\ \gamma|\mu\nu}_{\delta\epsilon|\rho}
         \bigg\}\end{split}\,,\\
&\begin{split}(i10),\quad 0=\bigg\{&+\eta^{\psi\gamma}\,t_{\sss \d\d u  h}(z)^{\ \: |\mu\nu}_{\delta\epsilon|}+
           t^{\psi}_{\sss \d\d u  h}(z)^{\ \gamma|\mu\nu}_{\delta\epsilon|}\bigg\}\end{split}\,.
\label{eq:b17.9}
\end{split}
\end{equation}
These identities hold among the C-number $2$-point distributions constructed in second order perturbation theory. Some of them 
have been explicitly checked by calculations, but there is no doubt about their validity, because in  Sec.~\ref{sec:loop2}, the 
condition~(\ref{eq:b17.2}) of perturbative gauge invariance to second order for loop graphs is proved. 

In the case of QG coupled to photon fields~\cite{gri5} and scalar matter fields~\cite{gri6}, these identities are less involved 
and from them we can derive easily the Slavnov--Ward identities for the $2$-point connected Green function with photon and
matter loop, respectively. We will return on these identities and their relation to the gravitational Slavnov--Ward identities
in Sec.~4.

\section{Two-Point Distribution for Graviton Self-Energy}\label{sec:twopoint}
\setcounter{equation}{0}

It is our aim in this section to apply the causal scheme to QG in order to calculate the $2$-point 
distribution $T_{2}(x,y)$ which describes the graviton self-energy contribution. 
We explain step by step how $T_{2}(x,y)$ has to be constructed 
according to the general rules of the  causal scheme~\cite{scha}.

The are two important pieces in the inductive calculation that we are going to carry out: 
the first one is  the calculation in momentum space  of the product of positive/negative parts of Jordan--Pauli 
distributions (see App.~1 and App.~2 for the technical details) and the second one is  the causal splitting procedure 
(see Sec.~\ref{sec:splitting}) according to the correct singular order (see Sec.~\ref{sec:singord}).

\subsection{Inductive Construction}\label{sec:inductive}

First of all, from the commutation rules~(\ref{eq:b7}) and~(\ref{eq:b15}) we compute the contractions between two field 
operators:
\begin{gather}
C\big\{ h^{\alpha\beta}(x)\ h^{\mu\nu}(y) \big\}:=\big[h^{\alpha\beta}(x)^{\sss (-)},h^{\mu\nu}(y)^{\sss (+)}\big]=-i\,
b^{\alpha\beta\mu\nu}\, D_{0}^{\sss (+)}(x-y)\,,\nonumber \\
C\big\{ u^{\mu}(x)\ \tilde{u}^{\nu}(y) \big\} :=\big\{ u^{\mu}(x)^{\sss (-)}, \tilde{u}^{\nu}(y)^{\sss (+)}\big\}=+i\,
\eta^{\mu\nu}\,D_{0}^{\sss (+)}(x-y)\,,\nonumber \\
C\big\{ \tilde{u}^{\mu}(x)\ {u}^{\nu}(y) \big\} :=\big\{ \tilde{u}^{\mu}(x)^{\sss (-)}, {u}^{\nu}(y)^{\sss (+)}\big\}=
-i\,\eta^{\mu\nu}\,D_{0}^{\sss (+)}(x-y)\,;
\label{eq:b18}
\end{gather}
where $(\pm)$ refers to the positive/negative frequency part of the corresponding quantity.

The first step in the construction of $T_{2}(x,y)$ consists in  calculating the auxiliary distributions
\begin{equation}
R'_{2}(x,y):= - T^{h+u}_{1}(y)\, T^{h+u}_{1}(x)\,,\quad A'_{2}(x,y):= - T^{h+u}_{1}(x)\, T^{h+u}_{1}(y)\,
\label{eq:b19}
\end{equation}
from these we form  the causal distribution
\begin{equation}
D_{2}(x,y):=R'_{2}(x,y)-A'_{2}(x,y)=\Big[T^{h+u}_{1}(x),T^{h+u}_{1}(y)\Big]\,.
\label{eq:b20}
\end{equation}
Causal means that the numerical part of $D_{2}(x,y)$ has support inside the light cone. Being $T_{1}^{h+u}(x)$ a normally 
ordered product, we have to carry  out all  possible contractions between the two factors in~(\ref{eq:b19})  using 
Wick's lemma. In this 
manner  $D_{2}(x,y)$ contains tree contributions or scattering graphs (only one contraction and four external legs), loop 
contributions (two contractions and two external legs) and vacuum graph contributions (three contractions and no external 
legs). Note that, due to the presence of normal ordering, tadpole diagrams do not appear in causal perturbation theory. 


\subsection{Example of the Calculation}\label{sec:example}

Let us illustrate how to construct $D_{2}(x,y)$ by explicitly working out an example. We take into account only the first 
term in the graviton coupling $T_{1}^{h}(x)$ so that,  from the $A^{'}_{2}(x,y)$-distribution, that can be decomposed as a sum 
of $25$ different contributions $\sum_{i,j=1}^{5}A'_{2}(x,y)^{\sss (i,j)}$, we pick up only the term 
$A'_{2}(x,y)^{\sss (1,1)}$, and in addition we carry out the loop generating double-contractions only between graviton 
fields that carry derivatives, so that for  $A'_{2}(x,y)^{\sss (1,1)}$ we obtain
\begin{multline}
A'_{2}(x,y)^{\sss (1,1)}=-\left(\frac{i\kappa}{2}\right)^{2}:\!\underbrace{h^{\alpha\beta}h^{\rho\sigma}_{\: ,\alpha}
h^{\rho\sigma}_{\:,\beta}}_{x}\!:\,  :\!\underbrace{h^{\mu\nu}h^{\gamma\delta}_{\: ,\mu}
h^{\gamma\delta}_{\:,\nu}}_{y}\!:\Big|_{\substack{\text{2 contractions}}}=\\
\begin{split}&=+\frac{\kappa^2}{4}\,:\!h^{\alpha\beta}(x)h^{\mu\nu}(y)\!:
\bigg[  + C\big\{ h^{\rho\sigma}(x)_{,\alpha}\,h^{\gamma\delta}(y)_{,\mu} \big\}\cdot
         C\big\{ h^{\rho\sigma}(x)_{,\beta} \,h^{\gamma\delta}(y)_{,\nu} \big\}+\\
      &\quad + C\big\{ h^{\rho\sigma}(x)_{,\alpha}\,h^{\gamma\delta}(y)_{,\nu} \big\}\cdot
         C\big\{ h^{\rho\sigma}(x)_{,\beta} \,h^{\gamma\delta}(y)_{,\mu} \big\} \bigg]+\text{other contractions}\,.
\label{eq:b22}
\end{split}
\end{multline}
Using the relations in~(\ref{eq:b18}), we find
\begin{equation}
\begin{split}
A'_{2}(x,y)^{\sss (1,1)}&=+\frac{\kappa^2}{4}\,:\!h^{\alpha\beta}(x)h^{\mu\nu}(y)\!:
\Big[+\big(-i\,b^{\rho\sigma\gamma\delta}\partial_{\alpha}^{x}\partial_{\mu}^{y}D_{0}^{\sss (+)}(x-y)\big)\\
    &\quad\times  \big(-i\,b^{\rho\sigma\gamma\delta}\partial_{\beta}^{x}\partial_{\nu}^{y}D_{0}^{\sss (+)}(x-y)\big) 
+\big(-i\,b^{\rho\sigma\gamma\delta}\partial_{\alpha}^{x}\partial_{\nu}^{y}D_{0}^{\sss (+)}(x-y)\big)\\
   &\quad\times   \big(-i\,b^{\rho\sigma\gamma\delta}\partial_{\beta}^{x}\partial_{\mu}^{y}D_{0}^{\sss (+)}(x-y)\big) \Big]+
\text{other contractions}\,.
\label{eq:b23}
\end{split}
\end{equation}
Since $b^{\rho\sigma\gamma\delta}b_{\rho\sigma\gamma\delta}=10$ and $\partial_{\mu}^{y}D_{0}^{\sss (+)}(x-y)=-
\partial_{\mu}^{x}D_{0}^{\sss (+)}(x-y)$ we obtain
\begin{equation}
A'_{2}(x,y)^{\sss (1,1)}=:\!h^{\alpha\beta}(x)h^{\mu\nu}(y)\!:\,
a'_{2}(x-y)_{\alpha\beta\mu\nu}^{\sss (1,1)}+\ldots\,,
\label{eq:b24}
\end{equation}
where
\begin{equation}
\begin{split}
a'_{2}(x-y)_{\alpha\beta\mu\nu}^{\sss (1,1)}&:=-\frac{5\,\kappa^2}{2}\,\Big( D^{\sss (+)}_{\alpha\mu|\beta\nu}(x-y)
+D^{\sss (+)}_{\alpha\nu|\beta\mu}(x-y)\Big)\,,\\
D^{\sss (+)}_{\alpha\mu|\beta\nu}(x-y)&:=\partial_{\alpha}^{x}\partial_{\mu}^{x}D_{0}^{\sss (+)}(x-y)\cdot
   \partial_{\beta}^{x}\partial_{\nu}^{x}D_{0}^{\sss (+)}(x-y)\,.
\label{eq:b25}
\end{split}
\end{equation}
The products between derivatives of Jordan--Pauli distributions are calculated in App.~1. Analogously, by taking into 
account that $D_{0}^{\sss (+)}(y-x)=-D_{0}^{\sss (-)}(x-y)$, we find
\begin{equation}
R'_{2}(x,y)^{\sss (1,1)}=:\!h^{\alpha\beta}(x)h^{\mu\nu}(y)\!:\,
r'_{2}(x-y)_{\alpha\beta\mu\nu}^{\sss (1,1)}+\ldots\,,
\label{eq:b26}
\end{equation}
where
\begin{equation}
\begin{split}
r'_{2}(x-y)_{\alpha\beta\mu\nu}^{\sss (1,1)}&:=-\frac{5\,\kappa^2}{2}\,\Big( D^{\sss (-)}_{\alpha\mu|\beta\nu}(x-y)
+D^{\sss (-)}_{\alpha\nu|\beta\mu}(x-y)\Big)\,,\\
D^{\sss (-)}_{\alpha\mu|\beta\nu}(x-y)&:=\partial_{\alpha}^{x}\partial_{\mu}^{x}D_{0}^{\sss (-)}(x-y)\cdot
   \partial_{\beta}^{x}\partial_{\nu}^{x}D_{0}^{\sss (-)}(x-y)\,.
\label{eq:b27}
\end{split}
\end{equation}
Therefore, according to Eq.~(\ref{eq:b20}), the $D_{2}(x,y)$-distribution has the form
\begin{gather}
D_{2}(x,y)^{\sss (1,1)}=:\!h^{\alpha\beta}(x)h^{\mu\nu}(y)\!:\,d_{2}(x-y)^{\sss (1,1)}_{\alpha\beta\mu\nu}+\ldots\,,
\nonumber \\
d_{2}(x-y)_{\alpha\beta\mu\nu}^{\sss (1,1)}=r'_{2}(x-y)_{\alpha\beta\mu\nu}^{\sss (1,1)}-
      a'_{2}(x-y)_{\alpha\beta\mu\nu}^{\sss (1,1)}\,.
\label{eq:b28}
\end{gather}

The most important property of $D_{2}(x,y)$ is causality, but only the numerical distribution $d_{2}(x-y)$ 
is responsible for this  property:
\begin{equation}
\hbox{supp}\big(d_{2}(z)\big)\subseteq \overline{ V^{+}(z)}\cup \overline{ V^{-}(z)}\,,
\quad\text{with}\ z:=x-y\,,
\label{eq:b29}
\end{equation}
(see below).
The products of Jordan--Pauli distributions appearing in~(\ref{eq:b28})  are easily expressed in 
momentum space, see App. 2, so that we obtain
\begin{gather}
\hat{a}'_{2}(p)^{\sss (1,1)}_{\alpha\beta\mu\nu}=-\hat{P}(p)^{\sss (4)}_{\alpha\beta\mu\nu}\,\Theta(p^2)\,
                           \Theta(+p^0)\,,\nonumber\\
\hat{r}'_{2}(p)^{\sss (1,1)}_{\alpha\beta\mu\nu}=-\hat{P}(p)^{\sss (4)}_{\alpha\beta\mu\nu}\,\Theta(p^2)\,
\Theta(-p^0)\,;
\label{eq:b31}
\end{gather}
and therefore
\begin{equation}
\hat{d}_{2}(p)^{\sss (1,1)}_{\alpha\beta\mu\nu}=\hat{P}(p)^{\sss (4)}_{\alpha\beta\mu\nu}\,\Theta(p^2)\,
\Big[\Theta(p^0 )
-\Theta(-p^0 )\big]=\hat{P}(p)_{\alpha\beta\mu\nu}\,\Theta(p^2)\,\mathrm{sgn}(p^0)\,,
\label{eq:b32}
\end{equation}
where $\hat{P}(p)^{\sss (4)}_{\alpha\beta\mu\nu}$ is a Lorentz covariant polynomial of degree four, this 
degree is given by the number of derivatives on the contracted lines. Causality is evident from the scalar 
distribution $\hat{d}(p):=\Theta(p^2)\mathrm{sgn}(p^0)$. For $z^2<0$, we may choose a reference frame in 
which $z^{\alpha}=(0,\bol{z})$, so that
\begin{equation}
d(z)=\frac{1}{(2\pi)^2}\int\!\!dp^0\,\mathrm{sgn}(p^{0}) \int\!\! d^{3}p\,\Theta(p_{0}^{2}-\bol{p}^2)\,
       e^{+i\,\bol{p}\cdot\bol{z}}=0\,,
\label{eq:b33}
\end{equation}
because of the signum-function in $p^{0}$. Therefore $d(z)$ vanishes 
outside the light cone, see Eq.~(\ref{eq:b29}).

\subsection{Causal \boldmath$D_{2}(x,y)$-Distribution for Graviton Self-Energy}\label{sec:d2distr}

The total $D_{2}(x,y)$-distribution for the graviton self-energy through a graviton loop is obtained by calculating the $25$ 
contributions coming from the graviton coupling $T_{1}^{h}$, not only the terms with two external graviton fields without 
derivatives, but also these with one or two derivatives. In addition, there are also $16$ contributions coming from the 
ghost-graviton coupling where one performs two ghost--anti-ghost contractions. Summing graviton loop and ghost--anti-ghost 
loop contributions we obtain~\footnote{the notation `$\cdot|\cdot$'  keeps track of the exact position of 
the indices}
\begin{equation}
\begin{split}
D_{2}(x,y)=&+:\!h^{\alpha\beta}(x)h^{\mu\nu}(y)\!:\,
                d^{\sss (4)}_{2}(x-y)^{\cdot|\cdot}_{\alpha\beta\mu\nu} +\\
  & +:\!h^{\alpha\beta}(x)_{,\gamma}h^{\mu\nu}(y)\!:\,
d^{\sss (3a)}_{2}(x-y)^{\gamma|\cdot}_{\alpha\beta\mu\nu}+\\
&+:\!h^{\alpha\beta}(x)h^{\mu\nu}(y)_{,\rho}\!:\,d^{\sss (3b)}_{2}(x-y)^{\cdot|\rho}_{\alpha\beta\mu\nu}+\\
&+:\!h^{\alpha\beta}(x)_{,\gamma}h^{\mu\nu}(y)_{,\rho}\!:\,
d^{\sss (2)}_{2}(x-y)^{\gamma|\rho}_{\alpha\beta\mu\nu} \,.
\label{eq:b34}
\end{split}
\end{equation}
The tensorial distributions have in momentum space the structure
\begin{equation}
\hat{d}^{\sss (i)}_{2}(p)^{\cdot|\cdot}_{\alpha\beta\mu\nu}=\hat{P}^{\sss (i)}(p)_{\alpha\beta\mu\nu}^{\cdot|\cdot}
\,\hat{d}(p)\,,\quad
i=4,3a,3b,2\,.
\label{eq:b35}
\end{equation}
The results for  the tensorial distributions, being too long, are not given here. See App.~4 for the explicit
form of the distributions appearing in Eq.~(\ref{eq:b34}). 
These will be used to calculate the graviton self-energy tensor in Sec.~\ref{sec:selfenergy}.

In order to obtain $T_{2}(x,y)$, we have to split the $D_{2}$-distribution into a retarded part, $R_{2}$, and an 
advanced part, $A_{2}$, with respect to the coincidence  point $z:=x-y=0$, so that 
$\hbox{supp}(R_{2}(z))\subseteq \overline{ V^{+}(z)}$ and $\hbox{supp}(A_{2}(z))\subseteq \overline{ V^{-}(z)}$. 
The correct treatment of this coincidence  point constitutes the key to control the UV behaviour of the $2$-point 
distribution. 

This splitting procedure affects only the numerical distributions $d_{2}^{\sss (i)}$ in~(\ref{eq:b34})
and must be accomplished according to the correct singular order $\omega(d_{2}^{\sss (i)})$ of the distribution. It
describes the behaviour of $d_{2}^{\sss (i)}(z)$ near the coincidence point $z=0$, or that  of $\hat{d}_{2}^{\sss (i)}(p)$ 
for $p\to\infty$. If $\omega(d_{2}^{\sss (i)}) < 0$, then the splitting is trivial. On the other side, if $\omega(d_{2}^{(i)})\ge 0$, 
then the splitting is non-trivial and non-unique: 
\begin{equation}
d_{2}^{\sss (i)}(x-y)_{\alpha\beta\mu\nu}^{\cdot|\cdot}\longrightarrow r_{2}^{\sss (i)}(x-y)_{\alpha\beta\mu\nu}^
{\cdot|\cdot}+\sum_{|a| =0}^{\omega(d_{2}^{\sss (i)})} \big\{ C_{a,i}\, D^{a}\big\}_{\alpha\beta\mu\nu}^{\cdot|\cdot}\, 
\delta^{\sss (4)}(x-y)\, ,
\label{eq:b36}
\end{equation}
and a retarded part $r_{2}^{\sss (i)}(x-y)$ is best obtained in momentum space by means of a dispersion-like integral, 
see Sec.~\ref{sec:splitting}, which, however, presents some difficulties in the massless case.
 
The $C_{a,i}$'s in Eq.~(\ref{eq:b36}) are undetermined but finite normalization constants and  $D^{a}$ is a partial 
differential operator acting on the local distribution $\delta^{\sss (4)}(x-y)$.
The second term on the right side of Eq.~(\ref{eq:b36}) represents therefore a  freedom in the normalization which is inherent to
the causal splitting of distributions and will be discussed in Sec.~\ref{sec:norm} by taking physical conditions into account. 

In the case of Eq.~(\ref{eq:b35}), we find from direct inspection of the distributions that
\begin{equation}
\omega(d_{2}^{\sss (4)})=4\,,\quad\omega(d_{2}^{\sss(3a)})=3\,,\quad\omega(d_{2}^{\sss(3b)})=3\,,
\quad\omega(d_{2}^{\sss(2)})=2\,.
\label{eq:b37}
\end{equation}
The singular order depends on the structure of the graph, namely on the number of derivatives acting on the contracted 
internal lines of the loop. For a precise formulation, see below.

The last step in the inductive construction of $T_{2}(x,y)$ consists in subtracting $R'_{2}(x,y)$ from $R_{2}(x,y)$, 
see Sec.~\ref{sec:splitting}. The singular order remains unchanged after distribution splitting: 
$\omega(d_{2})=\omega(r_{2})=\omega(t_{2})$.

\subsection{Singular Order in Quantum Gravity}\label{sec:singord}

Before undertaking the splitting of $D_{2}(x,y)$ according to Eq.~(\ref{eq:b36}), we give the formula for the singular 
order of arbitrary  $n$-point distributions in perturbative quantum gravity.

We consider in the $n$-th order of perturbation theory an arbitrary $n$-point distribution 
$T^{\sss G}_{n}(x_{1},\ldots,x_{n})$, 
appearing in Eq.~(\ref{eq:b1}), as a sum of normally ordered products of free field operators multiplied by numerical 
distributions
\begin{equation}
T^{\sss G}_{n}(x_{1},\ldots ,x_{n})
= :\!\prod_{j=1}^{n_h}h(x_{k_{j}})\,\prod_{i=1}^{n_{u}}u(x_{m_{i}})\,\prod_{l=1}^{n_{\tilde{u}}}
\tilde{u}(x_{n_{l}})\!:\,t^{\sss G}_{n}(x_{1},\ldots ,x_{n})\,.
\label{eq:b38}
\end{equation}
This $T_{n}^{\sss G}$ corresponds to a graph $G$ with $n_{h}$ external graviton lines , $n_{u}$ external ghost lines  
and $n_{\tilde{u}}$ external anti-ghost lines. The singular order of $G$ then reads
\begin{equation}
\omega(G)\le 4-n_{h}-n_{u}-n_{\tilde{u}}-d+n\,.
\label{eq:b39}
\end{equation}
Here $d$ is the number of derivatives on the external field operators in~(\ref{eq:b38}). The `$\le$' means that in 
certain cases the singular order is lowered by peculiar conditions, \emph{e.g.} by the equations of motions of the free 
fields.

The explicit presence of the order of perturbation theory renders the theory `non-normalizable', that is the theory has 
a weaker predictive power but it is still well-defined in the sense of UV finiteness.

We give some hints of the inductive proof of~(\ref{eq:b39})~\cite{scha},~\cite{ym2}. First of all, the 
assumption~(\ref{eq:b39}) must be verified for $n=1$: $\omega\big(T_{1}^{h+u}(x)\big)$ has to be zero (because of 
$\omega\big(\delta(x)\big)=0$), a result which is correctly given by~(\ref{eq:b39}) after direct inspection.

In the inductive construction of $T_{n}$ from the $T_{m}$'s, $m\le n-1$, we must consider tensor products of two 
distributions
\begin{equation}
T_{r,1}(x_{1},\ldots,x_{r})\,T_{s,2}(y_{1},\ldots,y_{s})\,,
\label{eq:b40}
\end{equation}
with known singular order $\omega(T_{r,1})=\omega_{1}\le 4-n_{h_{1}}-n_{u_{1}}-n_{\tilde{u}_{1}}-d_{1}+r$
and $\omega(T_{s,2})=\omega_{2}\le 4-n_{h_{2}}-n_{{u}_{2}}-n_{\tilde{u}_{2}}-d_{2}+s$. According to the inductive 
construction, this product has to be normally ordered giving origin to all possible contraction configurations. We assume that $l$ 
contractions arise during this process. Taking translation invariance into account the numerical distribution of the 
contracted expression is of the form
\begin{gather}
t_{1}(x_{1}-x_{r},\ldots,x_{r-1}-x_{r})\,\prod_{j=1}^{l}\partial^{a_{j}} D_{0}^{\sss (+)}(x_{r_{j}}-y_{s_{j}})\,
t_{2}(y_{1}-y_{s},\ldots,y_{s-1}-y_{s})=\nonumber\\
=\tilde{t}(\xi_{1},\ldots,\xi_{r-1},\eta_{1},\ldots,\eta_{s-1},\eta)\,,
\label{eq:b41}
\end{gather}
with $\xi_{j}:=x_{j}-x_{r}$, $\eta_{j}:=y_{j}-y_{s}$, $\eta:=x_{r}-y_{s}$, $a_{j}=0,1,2$ and $a=\sum_{j=1}^{l}a_{j}$. Then, 
using the distributional definition  of the singular order~\cite{scha},  we may conclude that  
\begin{equation}
\omega(\tilde{t})=\omega_{1}+\omega_{2}+2l-4+a\,.
\label{eq:b42}
\end{equation}
Inserting the expressions for $\omega_{1}$ and $\omega_{2}$ in Eq.~(\ref{eq:b42}), we get
\begin{equation}
\omega(\tilde{t})\le 4-(n_{h_{1}}+n_{h_{2}}+n_{u_{1}}+n_{u_{2}}
+n_{\tilde{u}_{1}}+n_{\tilde{u}_{2}}-2l)-(d_{1}+d_{2}-a) +(r+s)\,.
\label{eq:b43}
\end{equation}
The first bracket represents the number $n_{h}+n_{u}+n_{\tilde{u}}$ of external fields after $l$ contractions, the 
second bracket gives the number $d$ of derivatives remaining on these external fields, if the $l$ contractions carry $a$ 
derivatives. Since $r+s=n$, Eq.~(\ref{eq:b39}) is proved.

In the usual QFT formulation, Eq.~(\ref{eq:b39}) would imply that QG is `non-renormalizable', since $\omega(G)$ 
increases without bound for higher orders in the perturbative expansion. This means that there is a `proliferation' of 
divergences and of counterterms to compensate them.

The situation is different in causal perturbation theory: we are facing in this case  a `non-normalizable' theory,
\emph{i.e.} each of its diagrams  is finite due to the causal splitting  method, but the number of the free, 
undetermined and finite normalization constants in~(\ref{eq:b36}) increases with $n$. The question  is then  to find enough 
physical conditions or requirements to fix  this increasing ambiguity in the normalization.


\subsection{Splitting of the \boldmath$D_{2}(x,y)$-Distribution}\label{sec:splitting}

We now carry out the splitting of the distribution $D_{2}(x,y)$ in  Eq.~(\ref{eq:b34}). 

Let us consider for example the numerical tensorial distribution 
$\hat{d}_{2}^{\sss (4)}(p)^{\cdot|\cdot}_{\alpha\beta\mu\nu}$ which has
singular order four from Eq.~(\ref{eq:b37}) of from Eq.~(\ref{eq:b39}). Because of the decomposition~(\ref{eq:b35}),
it suffices to split the scalar distribution $\hat{d}(p)=\Theta(p^2)\mathrm{sgn}(p^0)$ with $\omega(\hat{d})=0$ and
then multiply the so obtained retarded part by the same tensor 
$\hat{P}^{\sss (4)}(p)^{\cdot|\cdot}_{\alpha\beta\mu\nu}$ as given by Eq.~(\ref{eq:b35}).

Usually, a  special retarded part in  Eq.~(\ref{eq:b36}), if it exists,  is given in momentum space by the dispersion 
integral
\begin{equation}
\hat{r}^{0}(p)=\frac{i}{2\pi}\int_{-\infty}^{\infty}\!\! dt \frac{\hat{d}(tp)}{\left( t-i0\right)^{\omega +1}
\left( 1-t+i0\right)}\,,\quad  p \in \overline{V^{+}} \, ;
\label{eq:b44}
\end{equation}
which is called `central splitting solution', because the subtraction point~\cite{scha} is the origin. But this formula 
cannot be used directly in the case of massless theories, because the integral is divergent. In order to circumvent this 
deficiency, we shift the original distribution $d(x)$:
\begin{equation}
d_{q}(x):=e^{i\,q\cdot x}\,d(x)\,,\quad   \hat{d}_{q}(p)=\hat{d}(p+q)\,,\quad q^2<0\, ,
\label{eq:b45}
\end{equation}
so that the central splitting solution $\hat{r}_{q}^{0}(p)$ of the shifted distribution exists~\cite{ym2}.

We cannot obtain the retarded part of the original distribution simply by letting $q\to 0$, so we take advantage 
of the local ambiguity in the splitting procedure and consider another retarded part of $\hat{d}_{q}(p)$, given 
by $\hat{r}_{q}(p)$. Since two retarded distributions differ only by local terms in configuration space, we obtain in 
momentum space for fixed $q$ that the difference reads
\begin{equation}
\hat{r}_{q}(p)-\hat{r}_{q}^{0}(p)=\hat{P}_{q}(p)\,,
\label{eq:b46}
\end{equation}
where $\hat{P}_{q}(p)$ is a $q$-dependent polynomial in $p$ of degree $\omega$. Then we construct $\hat{r}(p)$,
a retarded part of the original distribution $\hat{d}(p)$, from~(\ref{eq:b46}) by taking the limit
\begin{equation}
\hat{r}(p)=\lim_{q\to 0}\,\hat{r}_{q}(p)=\lim_{q\to 0}\,\big[\hat{r}_{q}^{0}(p)+\hat{P}_{q}(p)\big]\,.
\label{eq:b47}
\end{equation}
Here the addition of the $q$-dependent polynomial $\hat{P}_{q}(p)$ must be accomplished in such a way that the limit 
exists. This corresponds to a finite renormalization.

Using~(\ref{eq:b44}) with~(\ref{eq:b45}) in~(\ref{eq:b47}), we obtain for $p \in V^{+}$, with $q\to 0$ in such a way that 
$p-q \in V^{+}$, $q^2<0$:
\begin{equation}
\begin{split}
\hat{r}(p)&=\lim_{q\to 0}\Big[\frac{i}{2\pi}\int_{-\infty}^{+\infty}\!\! dt
               \frac{\hat{d}_{q}(tp)}{\big(t-i0\big)\big(1-t+i0\big)} +\hat{P}_{q}(p)\Big]\\
 &=\lim_{q\to 0}\Big[\frac{i}{2\pi}\int_{-\infty}^{+\infty}\!\! dt
               \frac{\hat{d}(tp+q)}{\big(t-i0\big)\big(1-t+i0\big)} +\hat{P}_{q}(p)\Big]\\
&=\lim_{q\to 0}\Big[\frac{i}{2\pi}\int_{-\infty}^{+\infty}\!\!
    \frac{dt}{\big(t-i0\big)\big(1-t+i0\big)}\,\Theta\big((tp+q)^2\big)\,\mathrm{sgn}(tp^0+q^0)+
    \hat{P}_{q}(p)\Big]\,.
\raisetag{2cm}\label{eq:b48}
\end{split}
\end{equation}
The zeros of $(tp+q)^2$ are $t_{1,2}=\frac{1}{p^2}\big(-p\cdot q\pm\sqrt{N}\big)$ with $N:=(p\cdot q)^2-p^2q^2$, so that
the integral in~(\ref{eq:b48}) may be simplified to 
\begin{multline}
\bigg\{\frac{-i}{2\pi}\int_{-\infty}^{t_{2}<0}\!\!dt +\frac{i}{2\pi}\int_{t_{1}>0}^{\infty}\!\!dt \bigg\}\,
\Big(\frac{1}{t}-P\frac{1}{t-1} -i\,\pi\,\delta(t-1)\Big)=\\
=\frac{i}{2\pi}\bigg[-i\,\pi+\log\left(\frac{p^2}{|q^2|}\right)+O(\sqrt{|q^2 |})\bigg]\,,
\label{eq:b50}
\end{multline}
and the limit reads
\begin{equation}
\hat{r}(p)=\lim_{q\to 0}\bigg[\frac{i}{2\pi}\Big( -i\,\pi+\log\left(\frac{p^2}{|q^2|}\right)
+O(\sqrt{ |q^2 |})\Big) +\hat{P}_{q}(p)\bigg]\,.
\label{eq:b51}
\end{equation}
Being $\omega=0$, we can add the polynomial $\hat{P}_{q}(p):=\frac{i}{2\pi}\log\left(|q^2|/M^2\right)$, where $M>0$ is an arbitrary mass
scale, so that we obtain a Lorentz invariant splitting solution
\begin{equation}
\hat{r}(p)=\frac{i}{2\pi}\left(\log\left( \frac{p^2}{M^2}\right) -i\,\pi\right)\,,\quad p \in V^{+}\,.
\label{eq:b52}
\end{equation}
By analytic continuation in $\mathbb{R}^{4}+iV^{+}$, we find for $p\in\mathbb{R}^{4}$
\begin{equation}
\hat{r}(p)^{an}=\frac{i}{2\pi}\log\left( \frac{-p^2 - i\,p^{0} 0}{M^2}\right)\,.
\label{eq:b53}
\end{equation}

As pointed out at the end of Sec.~\ref{sec:d2distr}, the $T_{2}(x,y)$-distribution is obtained from $R_{2}(x,y)$ by 
subtracting $R'_{2}(x,y)$. This subtraction affects only the scalar distributions. Since 
$\hat{r}'(p)=-\Theta(p^2)\Theta(-p^0)$, we obtain
\begin{equation}
\begin{split}
\hat{t}(p)&=\hat{r}(p)^{an}-\hat{r}'(p)\\
&=\frac{i}{2\pi}\Big(\log\left(\frac{|-p^2 |}{M^2}\right) -i\,\pi\,\mathrm{sgn}(p^0)\,\Theta(p^2 )\Big) +
 \Theta(p^2)\Theta(-p^0)\\
&=\frac{i}{2\pi}\Big(\log\left(\frac{|-p^2 |}{M^2}\right) -i\,\pi\,\Theta(p^2 )\Big)
=\frac{i}{2\pi}\log\left( \frac{-p^2 -i\,0}{M^2}\right)\,.
\label{eq:b54}
\end{split}
\end{equation}
The ambiguity in the  normalization  present in the splitting $D_{2}\to R_{2}+N_{2}$, Eq.~(\ref{eq:b36}), will be discussed in 
Sec.~\ref{sec:norm}.

\subsection{Graviton Self-Energy Tensor from the \boldmath$T_{2}(x,y)$-Distribution}\label{sec:selfenergy}

Gathering all the results of the previous sections, Eqs.~(\ref{eq:b34}),~(\ref{eq:b35}) and~(\ref{eq:b54}), we find that 
the $2$-point distribution that contributes to the graviton self-energy  reads
\begin{equation}
\begin{split}
T_{2}(x,y)=&+:\!h^{\alpha\beta}(x)h^{\mu\nu}(y)\!:\,
     t^{\sss (4)}_{2}(x-y)^{\cdot|\cdot}_{\alpha\beta\mu\nu} +\\
          & +:\!h^{\alpha\beta}(x)_{,\gamma}h^{\mu\nu}(y)\!:\,t^{\sss (3a)}_{2}(x-y)^{\gamma|\cdot}_{\alpha\beta\mu\nu}+\\
&+:\!h^{\alpha\beta}(x)h^{\mu\nu}(y)_{,\rho}\!:\,t^{\sss (3b)}_{2}(x-y)^{\cdot|\rho}_{\alpha\beta\mu\nu}+\\
&+:\!h^{\alpha\beta}(x)_{,\gamma}h^{\mu\nu}(y)_{,\rho}\!:\,t^{\sss (2)}_{2}(x-y)^{\gamma|\rho}_{\alpha\beta\mu\nu} \,,
\label{eq:b55}
\end{split}
\end{equation}
where the  tensorial distributions have in momentum space the structure
\begin{equation}
\hat{t}^{\sss (i)}_{2}(p)^{\cdot|\cdot}_{\alpha\beta\mu\nu}=\hat{P}^{\sss (i)}(p)_{\alpha\beta\mu\nu}^{\cdot|\cdot}
\,\hat{t}(p)\,,\quad i=4,3a,3b,2\,,
\label{eq:b56}
\end{equation}
where $\hat{t}(p)$ is given in Eq.~(\ref{eq:b54}) and the polynomials are those of Eq.~(\ref{eq:b35}).

The ${t}_{2}^{\sss (i)}$-distributions appearing in Eq.~(\ref{eq:b55}) have already been introduced
in Eq.~(\ref{eq:b17.1}): $t_{2}^{\sss (4)}=i {t}_{\sss hh}$, 
${t}_{2}^{\sss (3a)}=i {t}_{\sss \d h h}$,  ${t}_{2}^{\sss (3b)}=i {t}_{\sss  h\d h}$ and
${t}_{2}^{\sss (2)}=i {t}_{\sss \d  h\d h}$.

Since divergences in the adiabatic limit of Eq.~(\ref{eq:b1}) do not contribute, we can obtain from $T_{2}(x,y)$ by 
partial integration the graviton self-energy contribution
\begin{equation}
T_{2}(x,y)^{h\sss SE}=:\!h^{\alpha\beta}(x)h^{\mu\nu}(y)\!:\, i\, \Pi(x-y)_{\alpha\beta\mu\nu}\,.
\label{eq:b57}
\end{equation}
The main result of our calculation in second order causal perturbation theory is the graviton self-energy tensor 
$\Pi(x-y)_{\alpha\beta\mu\nu}$ which is given by the following combination of $t_{2}(x-y)$-distributions
\begin{equation}
\begin{split}
i\,\Pi(x-y)_{\alpha\beta\mu\nu}:= &+t_{2}^{\sss (4)}(x-y)^{\cdot|\cdot}_{\alpha\beta\mu\nu} 
                              -\d^{x}_{\gamma}t^{\sss (3a)}_{2}(x-y)^{\gamma|\cdot}_{\alpha\beta\mu\nu}+\\
&+\d_{\rho}^{x}t^{\sss (3b)}_{2}(x-y)^{\cdot|\rho}_{\alpha\beta\mu\nu}
-\d^{x}_{\gamma}\d_{\rho}^{x}t^{\sss (2)}_{2}(x-y)^{\gamma|\rho}_{\alpha\beta\mu\nu} \,,
\label{eq:b58}
\end{split}
\end{equation}
where we have carried the derivatives acting on the external fields in Eq.~(\ref{eq:b55}) on the corresponding  
$t_{2}(x-y)$-distributions. In momentum space, it reads
\begin{equation}
\begin{split}
\hat{\Pi}(p)_{\alpha\beta\mu\nu}&=\hat{\Pi}(p)^{\text{grav. loop}}_{\alpha\beta\mu\nu}
                      +\hat{\Pi}(p)^{\text{ghost loop}}_{\alpha\beta\mu\nu}=\\
&\begin{split}=\frac{\kappa ^2 \pi}{960 (2\pi)^5}\,
\bigg[
&-656\:  p^{\alpha}p^{\beta}p^{\mu}p^{\nu} -208\:  p^2\big( p^{\alpha}p^{\beta}\eta^{\mu\nu}+p^{\mu}p^{\nu}\eta^{\alpha\beta}\big)+\\
&+162\:  p^2 \big(p^{\alpha}p^{\mu}\eta^{\beta\nu}+p^{\alpha}p^{\nu}\eta^{\beta\mu}+p^{\beta}p^{\mu}\eta^{\alpha\nu}+
                   p^{\beta}p^{\nu}\eta^{\alpha\mu}\big)+ \\
&-162\:  p^4 \big(\eta^{\alpha\mu}\eta^{\beta\nu}+\eta^{\alpha\nu}\eta^{\beta\mu}\big)\\
&+118\:  p^4 \eta^{\alpha\beta}\eta^{\mu\nu}\bigg]\,\log\left(\frac{-p^2-i0}{M^2}\right)\,.
\end{split}
\label{eq:b59}
\end{split}
\end{equation}
Separate calculations for the graviton loop and ghost loop give the following contributions to the graviton self-energy 
tensor, respectively (see App.~4):
\begin{gather}
\hat{\Pi}(p)^{\text{grav. loop}}_{\alpha\beta\mu\nu}=\Xi\,\Big[-880,-260,+160,-170,+110\Big]\,
                         \log\left((-p^2-i0)/M^2 \right)\,,\nonumber\\
\hat{\Pi}(p)^{\text{ghost loop}}_{\alpha\beta\mu\nu}=\Xi\,\Big[+224,+52,+2,+8,+8\Big]\,\log\left((-p^2-i0)/M^2 \right)\,;
\label{eq:b60}
\end{gather}
where we have adopted  the convention of writing only the coefficients of the tensor according to the structure given in
Eq.~(\ref{eq:b59}) and $\Xi:=\kappa ^2 \pi / 960 (2\pi)^5$. 

Our result, numerical coefficients and logarithmic dependence on $p^2 /M^2$, agrees with the finite part of previous 
calculations~\cite{cap},~\cite{zaidi} obtained using {\emph{ad-hoc}} regularization schemes. As a consequence, the absence of UV 
divergences means that we do not need to add counterterms~\cite{hooft},~\cite{velt} involving four derivatives to the 
original Hilbert--Einstein Lagrangian in order to obtain UV finite radiative corrections to the graviton propagator. 
In our approach, all the expressions are cutoff-free and finite  at  each stage of the calculation due to the causal scheme.

\section{\hspace{-1mm}Gravitational Slavnov--Ward Identities and Perturbative Gauge Invariance}\label{sec:slav}
\setcounter{equation}{0}

\textsl{Gravitational Slavnov--Ward Identities}
\vspace{3mm}

\noindent
The gravitational Slavnov--Ward identities (SWI)~\cite{cap},~\cite{capSW},~\cite{zaidi},~\cite{delbo} are derived in standard 
quantum field theory from the connected Green functions. We construct the $2$-point connected Green function as
\begin{equation}
\hat{G}(p)_{\alpha\beta\mu\nu}^{\sss [2]}:=b_{\alpha\beta\gamma\delta}\,\hat{D}_{0}^{\sss F}(p)\,\hat{\Pi}(p)^{\gamma\delta\rho\sigma}\,
b_{\rho\sigma\mu\nu}\,\hat{D}_{0}^{\sss F}(p)\,,
\label{eq:b62}
\end{equation}
where $\hat{D}_{0}^{\sss F}(p)=(2\pi)^{-2} (-p^2-i0)^{-1}$. The two attached lines represent free graviton Feynman propagators. Then the 
gravitational SWI  reads
\begin{equation}
p^{\alpha}\,p^{\mu}\, \hat{G}(p)_{\alpha\beta\mu\nu}^{\sss [2]}=0\,.
\label{eq:b63}
\end{equation}
Since the tensorial structure of a general self-energy tensor may be characterized by the five coefficients 
$A,B,C,E,F$ in the standard representation $\hat{\Pi}(p)=\Xi\,\big[A,B,C,E,F\big]\,\log\left((-p^2-i0)/M^2\right)$ 
as in Eq.~(\ref{eq:b59}),  then the SWI are equivalent to the following relations
\begin{equation}
\frac{A}{4}-B+E+F=0\,,\quad C+E=0\,.
\label{eq:b66}
\end{equation}
These relations are satisfied by the coefficients of the self-energy tensor in Eq.~(\ref{eq:b59}), only if 
both graviton and ghost loops are taken into account. Therefore our result satisfies the SWI.
 
\vspace{5mm}
\noindent
\textsl{SWI and Other First Order Couplings}
\vspace{3mm}

\noindent
If we had chosen another ghost coupling instead of the one in Eq.~(\ref{eq:b14}), for example 
$T_{1}^{u}(x)=-\frac{i\,\kappa}{2}\big(+:\!\tilde{u}^{\mu} h^{\mu\nu}_{,\nu} u^{\rho}_{,\rho}\!:
+2\,:\!\tilde{u}^{\mu}_{,\rho} h^{\nu\rho} u^{\mu}_{,\nu}\!:\big)$~\cite{scho2}, then we would have violated
the SWI, because in this case the new graviton self-energy tensor through graviton and ghost  loop would have  had the form
\begin{equation}
\hat{\Pi}(p)^{\text{new}}_{\alpha\beta\mu\nu}=\Xi\,\big[-836,-238,+\frac{309}{2},-162,+118\big]\,
\log\left((-p^2-i0)/M^2\right)\,,
\label{eq:b67}
\end{equation}
so that it  would  have not satisfied  the SWI in Eq.~(\ref{eq:b66}).

Analogously, if we had disregarded the last two terms in the graviton coupling $T_{1}^{h}(x)$, 
Eq.~(\ref{eq:b6}), being a divergence due to the presence of two equal derivatives, then we would have violated the SWI, 
too, because in this case we would have  obtained a `reduced' graviton self-energy tensor through graviton loop of the form
\begin{equation}
\hat{\Pi}(p)^{\text{red.}}_{\alpha\beta\mu\nu}=\Xi\,\big[-880,+160,-108,+58,-62 \big]\,\log\left((-p^2-i0)/M^2\right)\,,
\label{eq:b68}
\end{equation}
so that the sum of graviton and ghost loop would have not satisfied  the SWI in Eq.~(\ref{eq:b66}). This happens although the 
difference between the two tensors can be written as a divergence
\begin{equation}
\hat{\Pi}(p)^{\text{grav. loop}}_{\alpha\beta\mu\nu}-\hat{\Pi}(p)^{\text{red.}}_{\alpha\beta\mu\nu}=\d_{\sigma}^{x}\Omega^{\sigma}(x-y)=
\text{divergence}\,,
\label{eq:b69}
\end{equation}
due to the vanishing of the coefficient proportional to $p^{\alpha}p^{\beta}p^{\mu}p^{\nu}$
and therefore this difference should not be physically relevant in the adiabatic limit $g\to 1$ of $S(g)$.

\vspace{3mm}
\noindent
\textsl{SWI and Perturbative Gauge Invariance}
\vspace{5mm}

\noindent
The SWI for QG coupled to photon fields~\cite{gri5} and matter fields~\cite{gri6} are equivalent to the transversality
of $\hat{G}(p)_{\alpha\beta\mu\nu}^{\sss [2]}$, namely $p^{\alpha}\hat{G}(p)_{\alpha\beta\mu\nu}^{\sss [2]}$. In the case
of self-coupled and ghost-coupled gravitons, the complexity of the gauge structure implies that only the weaker 
condition~(\ref{eq:b63}) can be satisfied by the graviton self-energy tensor, which is not transverse.

Perturbative gauge invariance to second order for loop graphs can be  formulated by means of the identities
$(i1)$,\ldots, $(i10)$  in Eq.~(\ref{eq:b17.9}). Note that these identities can be easily generalized to the  $n$-th order
for graphs with two external operators (`legs'). The identities will imply a relation among the graviton self-energy tensor
\begin{equation}
\Pi(p)^{\alpha\beta\mu\nu}=t_{\sss hh}(p)^{\alpha\beta\mu\nu} +i\,p_{\epsilon}\,t_{\sss \d h h}(p)^{\alpha\beta\mu\nu}_{\epsilon|}
-i\,p_{\rho}\,t_{\sss h\d  h}(p)^{\alpha\beta\mu\nu}_{\ |\rho}
+p_{\epsilon}p_{\rho}\,t_{\sss \d h\d h}(p)^{\alpha\beta\mu\nu}_{\epsilon|\rho}\,,
\label{eq:b69.1}
\end{equation}
the ghost self-energy tensors $t_{\sss u\d\tilde{u}}(p)$ and $t_{\sss \d u\d\tilde{u}}(p)$ and distributions coming from the $Q$-vertex,
namely $t_{\sss uh}$, $t_{\sss u\d h}$ and others  with one graviton and one ghost as external legs. We work in momentum space, but we 
skip the hat on the distributions.

From the structure of the first four identities in~(\ref{eq:b17.9}), it follows straightforwardly that
\begin{equation}
\begin{split}
b^{\gamma\delta\alpha\beta}\,&\Pi(p)^{\alpha\beta\mu\nu}=\frac{1}{2}\,\Big[ +  t^{ \delta  }_{\sss uh   }(p)^{\gamma|\mu\nu }   
+\eta^{\gamma\delta}\,t_{\sss uh   }(p)^{|\mu\nu }  -i\,p_{\tau}\, t^{\delta}_{\sss u\d h   }(p)^{\gamma|\mu\nu }_{\ |\tau   } +\\ 
&-i\,\eta^{\gamma\delta}\, p_{\tau}\, t_{\sss u\d h   }(p)^{|\mu\nu }_{ |\tau  } +\gamma \leftrightarrow \delta \Big]
+\frac{1}{4}\,\Big[+ p^2\, l_{\sss \d u\d h }(p)^{\gamma|\mu\nu}_{\delta|} +\\
&+p_{\epsilon} p_{\lambda}\,  l^{ \lambda  }_{\sss  \d u \d h}(p)^{\gamma|\mu\nu }_{\delta|\epsilon } 
+p_{\epsilon} p_{\nu}\,\tilde{l}_{\sss \d u \d h}(p)^{\gamma|\mu}_{\delta|\epsilon} +\gamma \leftrightarrow \delta+ 
\mu\leftrightarrow\nu\Big]\,.
\raisetag{15mm}\label{eq:b69.2}
\end{split}
\end{equation}
Here we have used $(i1)$,\ldots, $(i4)$ to express the graviton self-energy distributions by means of the $Q$-vertex distributions and
also $(i9)$, $(i10)$ to simplify the expression. Symmetrization of the left side is indeed necessary, because the right side is symmetric.
With $(i6)$, we can write the second bracket as
\begin{equation}
\frac{p_{\epsilon}p_{\tau}}{4}\,\Big[+\eta^{\nu\tau}\,t_{\sss \d u \d \tilde{u}}(p)^{\gamma|\mu}_{\delta|\epsilon}+
\mu\leftrightarrow\nu+\gamma\leftrightarrow \delta\Big]\,.
\label{eq:b69.3}
\end{equation}
The connected $2$-point Green function has a second $b$-tensor attached on the right, therefore we compute
\begin{equation}
\begin{split}
&b^{\gamma\delta\alpha\beta}\,\Pi(p)^{\alpha\beta\mu\nu}\,b^{\mu\nu\rho\sigma}=\frac{1}{2}\,
\bigg[\frac{1}{2}\, t_{\sss uh  }^{\delta }(p)^{\gamma|\rho\sigma}+\frac{1}{2}\, t_{\sss uh  }^{\delta }(p)^{\gamma|\sigma\rho}
              -\frac{1}{2}\, t_{\sss uh  }^{\delta }(p)^{\gamma|\mu\mu  }\,\eta^{\rho\sigma}+\gamma \leftrightarrow \delta   \\
&+\frac{1}{2}\,\eta^{\gamma\delta}\, t_{\sss uh}(p)^{|\rho\sigma} +\frac{1}{2}\,\eta^{\gamma\delta}\, t_{\sss uh}(p)^{|\sigma\rho}
     -\frac{1}{2}\,\eta^{\gamma\delta}\, t_{\sss uh}(p)^{|\mu\mu}\,\eta^{\rho\sigma}+\gamma \leftrightarrow \delta  +    \\
&-\frac{i}{2}\,p_{\epsilon}\,t_{\sss u\d h}^{\delta}(p)^{\gamma|\rho\sigma}_{\ |\epsilon} 
    -\frac{i}{2}\,p_{\epsilon}\,t_{\sss u\d h}^{\delta}(p)^{\gamma|\sigma\rho}_{\ |\epsilon} 
+\frac{i}{2}\,p_{\epsilon}\,t_{\sss u\d h}^{\delta}(p)^{\gamma|\mu\mu}_{\ |\epsilon}\,\eta^{\rho\sigma}+\gamma\leftrightarrow\delta+\\
&-\frac{i}{2}\,\eta^{\gamma\delta}\,p_{\epsilon}\,t_{\sss u\d h}(p)^{\gamma|\rho\sigma}_{\ |\epsilon} 
-\frac{i}{2}\,\eta^{\gamma\delta}\,p_{\epsilon}\,t_{\sss u\d h}(p)^{\gamma|\sigma\rho}_{\ |\epsilon} 
+\frac{i}{2}\,\eta^{\gamma\delta}\,p_{\epsilon}\,t_{\sss u\d h}(p)^{\gamma|\mu\mu}_{\ |\epsilon}\eta^{\rho\sigma} 
+\gamma \leftrightarrow \delta    \\
&+\frac{p_{\epsilon}p_{\tau}}{4}\,\Big[ \big(\eta^{\nu\tau} \,t_{\sss \d u \d \tilde{u}}(p)^{\gamma|\mu}_{\delta|\epsilon}+
     \mu\leftrightarrow\nu +\gamma\leftrightarrow \delta\big) 2\, b^{\mu\nu\rho\sigma}\Big]\bigg]\,.
\label{eq:b69.4}
\end{split}
\end{equation}
We apply now $p_{\gamma}$ to the above expression. This enables us to use other identities $\big((i7)$  and $(i8)\big)$, so that 
after a little work we get
\begin{equation}
\begin{split}
p_{\gamma}\,\big(b^{\gamma\delta\alpha\beta}\,\Pi(p)^{\alpha\beta\mu\nu}\,b^{\mu\nu\rho\sigma}\big)=
\frac{1}{2}\,\bigg[
&+p_{\gamma}\, t_{\sss uh }^{\delta  }(p)^{ \gamma|\rho\sigma} 
    -\frac{1}{2}\,p_{\gamma}\, t_{\sss uh  }^{\delta  }(p)^{\gamma|\mu\mu}\,\eta^{\rho\sigma}\\
&+p_{\delta}\, t_{\sss uh    }(p)^{|\rho\sigma   }
    -\frac{1}{2}\,p_{\delta}\, t_{\sss uh   }(p)^{|\mu\mu}\,\eta^{\rho\sigma}\\
&-{i}\,p_{\gamma} p_{\epsilon}\, t_{\sss u\d h  }^{\delta  }(p)^{ \gamma|\rho\sigma  }_{\ | \epsilon }
         +\frac{i}{2}\, p_{\gamma}p_{\epsilon}\,  t_{\sss u\d h  }^{\delta  }(p)^{\gamma|\mu\mu}_{\ |\epsilon}\,\eta^{\rho\sigma}\\
&-{i}\,p_{\delta} p_{\epsilon}\, t_{\sss u\d h  }(p)^{|\rho\sigma  }_{\, | \epsilon }
         +\frac{i}{2}\, p_{\delta} p_{\epsilon}\,  t_{\sss u\d h  }(p)^{|\mu\mu}_{\, |\epsilon}\,\eta^{\rho\sigma}\\
+\frac{p_{\epsilon}p_{\tau}p_{\gamma}}{4}\,\Big[\big(\eta^{\nu\tau} \,t_{\sss \d u \d \tilde{u}}(p)^{\gamma|\mu}_{\delta|\epsilon}
    & +\mu\leftrightarrow\nu +\gamma\leftrightarrow \delta\big) 2\, b^{\mu\nu\rho\sigma}\Big]+\\
+\frac{i p_{\epsilon}}{2}\,\Big[ 
-p_{\epsilon}\, l_{\sss u\d h}(p)^{\delta|\rho\sigma  }
&-p_{\epsilon}\, l_{\sss u\d h}(p)^{\delta|\sigma\rho  }
-p_{\lambda}\, l_{\sss u\d h}^{\lambda}(p)^{\delta|\rho\sigma  }_{\ |\epsilon}\\
-p_{\lambda}\, l_{\sss u\d h}^{\lambda}(p)^{\delta|\sigma\rho  }_{\ |\epsilon}
&-p_{\sigma}\, \hat{l}_{\sss u\d h}(p)^{\delta|\rho }_{\ |\epsilon}
-p_{\rho}\, \hat{l}_{\sss u\d h}(p)^{\delta|\sigma }_{\ |\epsilon}\\
+\eta^{\rho\sigma}\,\Big( p_{\epsilon}\, l_{\sss u\d h}(p)^{\delta|\mu\mu  }
 & +p_{\lambda}\, l_{\sss u\d h}^{\lambda}(p)^{\delta|\mu\mu }_{\ |\epsilon}
+p_{\lambda}\, \hat{l}_{\sss u\d h}(p)^{\delta|\lambda }_{\ |\epsilon}\Big)\Big]\bigg]\,.
\label{eq:b69.5}
\end{split}
\end{equation}
Using now the identity $(i5)$, we can express the $l_{\sss u\d h}$-distributions by means of the $t_{\sss u\d \tilde{u}}$-distributions:
\begin{equation}
\begin{split}
p_{\gamma}\,\big(b^{\gamma\delta\alpha\beta}\,\Pi(p)^{\alpha\beta\mu\nu}\,b^{\mu\nu\rho\sigma}\big)=
\frac{1}{2}\,\bigg[
&+p_{\gamma}\, t_{\sss uh }^{\delta  }(p)^{ \gamma|\rho\sigma} 
    -\frac{1}{2}\,p_{\gamma}\, t_{\sss uh  }^{\delta  }(p)^{\gamma|\mu\mu}\,\eta^{\rho\sigma}\\
&+p_{\delta}\, t_{\sss uh    }(p)^{|\rho\sigma   }
    -\frac{1}{2}\,p_{\delta}\, t_{\sss uh   }(p)^{|\mu\mu}\,\eta^{\rho\sigma}\\
&-{i}\,p_{\gamma} p_{\epsilon}\,t_{\sss u\d h  }^{\delta  }(p)^{ \gamma|\rho\sigma  }_{\ | \epsilon }
         +\frac{i}{2}\, p_{\gamma}p_{\epsilon}\,  t_{\sss u\d h  }^{\delta  }(p)^{\gamma|\mu\mu}_{\ |\epsilon}\,\eta^{\rho\sigma}\\
&-{i}\,p_{\delta} p_{\epsilon}\, t_{\sss u\d h  }(p)^{|\rho\sigma  }_{| \epsilon }
         +\frac{i}{2}\, p_{\delta} p_{\epsilon} \, t_{\sss u\d h  }(p)^{|\mu\mu}_{|\epsilon}\,\eta^{\rho\sigma}\\
+\frac{p_{\epsilon}p_{\gamma}}{2}\,\Big[  +p_{\sigma}\, t_{\sss \d u \d \tilde{u}}(p)^{\gamma|\rho}_{\delta|\epsilon}                
                                & +p_{\rho}\, t_{\sss \d u \d \tilde{u}}(p)^{\gamma|\sigma}_{\delta|\epsilon}    
                                 -p_{\nu}\, t_{\sss \d u \d \tilde{u}}(p)^{\gamma|\nu}_{\delta|\epsilon}\,\eta^{\rho\sigma}
                                      +\gamma\leftrightarrow \delta   \Big]+\\   
+\frac{i p_{\epsilon}}{2}\,\Big[ -p_{\sigma}\, t_{\sss u\d \tilde{u}}(p)^{\delta|\rho}_{\ |\epsilon}
                      & -p_{\rho}\, t_{\sss u\d \tilde{u}}(p)^{\delta|\sigma}_{\ |\epsilon} 
                        +p_{\nu}\, t_{\sss u\d \tilde{u}}(p)^{\delta|\nu}_{\ |\epsilon}\, \eta^{\rho\sigma}
\Big]\bigg]\,.
\label{eq:b69.6}
\end{split}
\end{equation}
This expression is the desired identity that involves the graviton self-energy tensor $\Pi(p)^{\alpha\beta\mu\nu}$, the ghost self-energy
tensors $t_{\sss \d u \d \tilde{u}}(p)$ and $ t_{\sss u\d \tilde{u}}(p)$ and the distributions coming from the $Q$-vertex. Schematically:
\begin{equation}
p_{\gamma}\,\big(b\,\Pi(p)\,b\big)^{\gamma\delta\rho\sigma}= X(t_{\sss uh}(p), t_{\sss u\d h}(p),\ldots)^{\delta,\rho\sigma}
+\big( t_{\sss \d u \d \tilde{u}}(p)+ t_{\sss u\d \tilde{u}}(p)       \big)^{\delta,\rho\sigma}\,,
\label{eq:b69.7}
\end{equation}
where $X^{\delta,\rho\sigma}$ represents the first eight distributions in~(\ref{eq:b69.6}). In QG, 
$(b\,\Pi(p)\,b)$ is not transversal, because the right side of~(\ref{eq:b69.7})
does not vanish. Note that the remaining $Q$-vertex distributions cannot be eliminated from~(\ref{eq:b69.7}) as already noticed 
in~\cite{capSW}, although in a very different approach to gauge invariance in QG.
By multiplying~(\ref{eq:b69.6}) with $p_{\rho}$, we obtain
\begin{equation}
\begin{split}
p_{\rho}\, p_{\gamma}\,\big(b^{\gamma\delta\alpha\beta}\,\Pi(p)^{\alpha\beta\mu\nu}\,b^{\mu\nu\rho\sigma}\big)=
\frac{1}{2}\,\bigg[
&+p_{\gamma}p_{\rho}\, t_{\sss uh }^{\delta  }(p)^{ \gamma|\rho\sigma} 
    -\frac{1}{2}\,p_{\gamma}p_{\sigma}\, t_{\sss uh  }^{\delta  }(p)^{\gamma|\mu\mu}\\
&+p_{\delta}p_{\rho}\, t_{\sss uh    }(p)^{|\rho\sigma   }
    -\frac{1}{2}\,p_{\delta}p_{\sigma}\, t_{\sss uh   }(p)^{|\mu\mu}\\
&-{i}\,p_{\rho}p_{\gamma} p_{\epsilon}\, t_{\sss u\d h  }^{\delta  }(p)^{ \gamma|\rho\sigma  }_{\ | \epsilon }
         +\frac{i}{2}\, p_{\sigma} p_{\gamma}p_{\epsilon} \, t_{\sss u\d h  }^{\delta  }(p)^{\gamma|\mu\mu}_{\ |\epsilon}\\
&-{i}\,p_{\delta} p_{\epsilon}\, t_{\sss u\d h  }(p)^{|\rho\sigma  }_{| \epsilon }
         +\frac{i}{2}\, p_{\delta} p_{\epsilon}\,  t_{\sss u\d h  }(p)^{|\mu\mu}_{|\epsilon}\\
+\frac{p_{\epsilon}p_{\gamma}p^2}{2}\,\Big[  t_{\sss \d u\d\tilde{u}}(p)^{\gamma|\sigma}_{\delta|\epsilon} &+
        t_{\sss \d u\d\tilde{u}}(p)^{\delta|\sigma}_{\gamma|\epsilon}      \Big]
 - \frac{i p_{\epsilon}p^2}{2}\,\Big[ t_{\sss u\d \tilde{u}}(p)^{\delta|\sigma}_{\ |\epsilon}
\Big]\bigg]\,.
\label{eq:b69.8}
\end{split}
\end{equation}
This equation does not \emph{a priori} imply the SWI~(\ref{eq:b63}). We can summarize Eq.~(\ref{eq:b69.8}) as
\begin{equation}
p_{\rho}\, p_{\gamma}\,\big(b\,\Pi(p)\,b\big)^{\gamma\delta\rho\sigma}=
p_{\rho}\,X(t_{\sss uh}(p), t_{\sss u\d h}(p),\ldots)^{\delta,\rho\sigma}
+p_{\rho}\,\big( t_{\sss \d u \d \tilde{u}}(p)+ t_{\sss u\d \tilde{u}}(p)\big)^{\delta,\rho\sigma}\,.
\label{eq:b69.9}
\end{equation}
With the explicit results for 
$t_{\sss \d u \d \tilde{u}}(p)$ and $ t_{\sss u\d \tilde{u}}(p)$ obtained in Sec.~7.1, we find that there must be a compensation
between $Q$-vertex contributions  $p_{\rho}X^{\delta,\rho\sigma}$  and ghost self-energy contributions so that the right side 
of~(\ref{eq:b69.9}) vanishes  identically, 
because this has already been verified by explicitly  calculating the left side of~(\ref{eq:b69.8}). This may also explain  why in 
the two cases with different $T_{1}^{u}$ or different $T_{1}^{h}$, investigated above, the SWI of Eq.~(\ref{eq:b63}) or 
Eq.~(\ref{eq:b66}) are not fulfilled: Eq.~(\ref{eq:b69.9}) still holds true, therefore the theory is still gauge invariant to second
order, but  both sides do not vanish identically.

\section{Fixing of the  Freedom in the Normalization of  the Self-Energy Tensor}\label{sec:norm}
\setcounter{equation}{0}

As seen in Sec.~\ref{sec:singord}, in causal perturbation theory the problem of eliminating infinitely many UV divergent
expressions is changed into the issue of fixing an increasing number of free undetermined local normalization terms that
arise as a consequence of distribution splitting in each order of perturbation theory.

\subsection{Normalization Terms of \boldmath$T_{2}(x,y)$}\label{sec:n2}

For simplicity, we consider only the  freedom in the normalization of  Eq.~(\ref{eq:b57}) instead of Eq.~(\ref{eq:b55}). Since 
the singular order is four, local normalization terms $N_{2}(x,y)\sim\delta^{\sss (4)}(x-y)$ of singular order 
$0,\ldots,4$  appear during the process of distribution splitting and in momentum space they can be written as
\begin{gather}
N_{2}(x,y)^{h\sss SE}=:\!h^{\alpha\beta}(x)h^{\mu\nu}(y)\!:\,i\,N(\d_{x},\d_{y})_{\alpha\beta\mu\nu}\,
\delta^{\sss (4)}(x-y)\,,\nonumber \\
\hat{N}(p)_{\alpha\beta\mu\nu}=\Big( \hat{N}^{\sss (0)}+\hat{N}^{\sss (2)}
+\hat{N}^{\sss (4)}\Big)(p)_{\alpha\beta\mu\nu}\,,
\label{eq:b70}
\end{gather}
where the odd terms are excluded owing to parity. $\hat{N}(p)^{\sss (i)}_{\alpha\beta\mu\nu}$ is a polynomial in $p$ of 
degree $i$ with  $i=0,2,4$. We assume in addition that only scalar constants should be considered, because vector-valued 
or tensor-valued constants may enter in conflict with Lorentz covariance. Therefore we make the following ansatz taking 
also the symmetries of $\Pi(x-y)_{\alpha\beta\mu\nu}$ into account
\begin{equation}
\begin{split}
&\hat{N}(p)^{\sss (0)}_{\alpha\beta\mu\nu}=\Xi\,\Big[+c_{1}\,\big(\eta_{\alpha\mu}\eta_{\beta\nu}+
        \eta_{\alpha\nu}\eta_{\beta\mu}\big)+c_{2}\,\eta_{\alpha\beta}\eta_{\mu\nu} \Big]\,,\\
&\begin{split}\hat{N}(p)^{\sss (2)}_{\alpha\beta\mu\nu}=\Xi\,\Big[&+c_{3}\,\big( p_{\alpha}p_{\beta}\eta_{\mu\nu}
   +p_{\mu}p_{\nu}\eta_{\alpha\beta}\big)+\\
 & +c_{4}\,\big( p_{\alpha}p_{\mu}\eta_{\beta\nu}+ p_{\alpha}p_{\nu}\eta_{\beta\mu}+p_{\beta}p_{\mu}\eta_{\alpha\nu}+
   p_{\beta}p_{\nu}\eta_{\alpha\mu}\big)+\\
 & +c_{5}\,p^{2}\,\big(\eta_{\alpha\mu}\eta_{\beta\nu}+\eta_{\alpha\nu}\eta_{\beta\mu}\big)
   +c_{6}\,p^{2}\,\eta_{\alpha\beta}\eta_{\mu\nu}\Big]\,,
\end{split}
\\
&\hat{N}(p)^{\sss (4)}_{\alpha\beta\mu\nu}=\Xi\,\Big[c_{7},c_{8},c_{9},c_{10},c_{11}\Big]\,.
\label{eq:b71}
\end{split}
\end{equation}
$c_{1},\ldots,c_{11}$ are undetermined real numbers. Requiring the SWI to hold, we can reduce the normalization 
polynomials  to
\begin{equation}
\begin{split}
\hat{N}(p)^{\sss (0)}_{\alpha\beta\mu\nu}&=0\,,\\
\hat{N}(p)^{\sss (2)}_{\alpha\beta\mu\nu}&=\Xi\,\Big[0,c_{5}+c_{6},-c_{5},c_{5},c_{6}\Big]\,p^{-2}\,,\\
\hat{N}(p)^{\sss (4)}_{\alpha\beta\mu\nu}&=\Xi\,\Big[c_{7},c_{10}+c_{11}+\frac{c_{7}}{4},-c_{10},c_{10},c_{11}\Big]\,;
\label{eq:b72}
\end{split}
\end{equation}
in such a way that only five undetermined coefficients remain to be fixed. The self-energy tensor supplemented 
by the normalization terms then reads
\begin{equation}
\hat{\Pi}(p)^{\sss N}_{\alpha\beta\mu\nu}=\hat{\Pi}(p)_{\alpha\beta\mu\nu}+\hat{N}(p)^{\sss (2)}_{\alpha\beta\mu\nu}+
\hat{N}(p)^{\sss (4)}_{\alpha\beta\mu\nu}\,.
\label{eq:b73}
\end{equation}

\subsection{Total Graviton Propagator and Mass and Coupling Constant Normalization}\label{sec:aufsumm}

The task of eliminating the remaining freedom in Eq.~(\ref{eq:b73}) can be accomplished in our case by considering the 
total graviton  propagator as a sum of the free graviton Feynman  propagator~\cite{gri3}
\begin{equation}
\langle\Omega| T\big\{h^{\alpha\beta}(x)h^{\mu\nu}(y)\big\}|\Omega \rangle =-i\, b^{\alpha\beta\mu\nu}\, 
     D^{\sss{F}}_{0}(x-y)\,,
\label{eq:b74}
\end{equation}
from Eq.~(\ref{eq:b7}) and the contraction~(\ref{eq:b18}), with an increasing number of self-energy insertions. 
In momentum space we therefore obtain the series
\begin{equation}
\begin{split}
\hat{D}(p)_{tot}^{\alpha\beta\mu\nu}&=+b^{\alpha\beta\mu\nu}\,\hat{D}^{\sss F}_{0}(p)   +\hat{D}^{\sss F}_{0}(p)\,
b^{\alpha\beta\gamma\delta}\,\tilde{\Pi}(p)^{\sss N}_{\gamma\delta\rho\sigma}\,b^{\rho\sigma\mu\nu}\,
\hat{D}^{\sss F}_{0}(p)+\\
&\quad+b^{\alpha\beta\gamma\delta}\,\hat{D}^{\sss F}_{0}(p)\,\tilde{\Pi}(p)^{\sss N}_{\gamma\delta\rho\sigma}\,
b^{\rho\sigma\epsilon\tau} \,\hat{D}^{\sss F}_{0}(p)\,\tilde{\Pi}(p)^{\sss N}_{\epsilon\tau\lambda\kappa}\,
b^{\lambda\kappa\mu\nu}\,\hat{D}^{\sss F}_{0}(p)+\ldots\\
&=\hat{D}^{\sss F}_{0}(p)\,\Big[b^{\alpha\beta\mu\nu}+b^{\alpha\beta\rho\sigma}\,
\tilde{\Pi}(p)^{\sss N}_{\rho\sigma\gamma\delta}\,\hat{D}(p)_{tot}^{\gamma\delta\mu\nu}\Big]\,,
\label{eq:b75}
\end{split}
\end{equation}
where $\tilde{\Pi}(p)^{\sss N}_{\alpha\beta\mu\nu}:=(2\pi)^4\,\hat{\Pi}(p)^{\sss N}_{\alpha\beta\mu\nu}$. After 
multiplying with $\big(\hat{D}^{\sss F}_{0}(p)\big)^{-1}$ and with $b_{\alpha\beta\epsilon\tau}$, we find
\begin{equation}
\Big[\big(\hat{D}^{\sss F}_{0}(p)\big)^{-1}\,b_{\epsilon\tau}^{\quad\gamma\delta}-l_{\epsilon\tau\rho\sigma}\,
\tilde{\Pi}(p)_{\sss N}^{\rho\sigma\gamma\delta}\Big]\cdot \hat{D}(p)^{tot}_{\gamma\delta\mu\nu}=l_{\epsilon\tau\mu\nu}\,.
\label{eq:b76}
\end{equation}
Since  $b_{\alpha\beta\mu\nu}b^{\mu\nu}_{\quad\rho\sigma}=l_{\alpha\beta\rho\sigma}:=\big(\eta_{\alpha\mu}\eta_{\beta\nu}+\eta_{\alpha\nu}\eta_{\beta\mu}\big)/2$ represents
the unity for rank-4 tensors, from Eq.~(\ref{eq:b76}) we can derive the form of the inverse of the total propagator:
\begin{equation}
\begin{split}
\big(\hat{D}(p)_{tot}^{-1}\big)_{\alpha\beta\mu\nu}&=b_{\alpha\beta\mu\nu}\,\big(\hat{D}^{\sss F}_{0}(p)\big)^{-1}-
\tilde{\Pi}(p)^{\sss N}_{\alpha\beta\mu\nu}\\
&=(2\pi)^{2}\,\Big[b_{\alpha\beta\mu\nu}\,(-p^2 -i0)-(2\pi)^{2}\,\hat{\Pi}(p)^{\sss N}_{\alpha\beta\mu\nu}\Big]\,.
\label{eq:b77}
\end{split}
\end{equation}

Our aim is to impose mass and coupling constant normalization conditions on the total graviton propagator,
therefore we have to invert the expression in~(\ref{eq:b77}). For this purpose we express the inverse of the total 
graviton propagator in the `projection operator' $Q(p)_{\alpha\beta\mu\nu}^{\sss (i)}$   basis, introduced in App.~3 and find
\begin{multline}
\big(\hat{D}(p)_{tot}^{-1}\big)_{\alpha\beta\mu\nu}=(2\pi)^{2}\,\sum_{i=1}^{6} Q(p)^{\sss (i)}_{\alpha\beta\mu\nu}\,\\
\times\Big\{+x_{i}^{b}\,(p^{2}-i0)-(2\pi)^{2}\,\Xi\,x_{i}^{\pi}\,\log\left((-p^2-i0)/M^2\right)\Big\}\,,
\label{eq:b78}
\end{multline}
with
\begin{equation}
\begin{split}
b_{\alpha\beta\mu\nu}&=\sum_{i=1}^{6} Q(p)^{\sss (i)}_{\alpha\beta\mu\nu}\,x_{i}^{b}\,(-p^2-i0)\,,\\
\hat{\Pi}(p)^{\sss N}_{\alpha\beta\mu\nu}&=\sum_{i=1}^{6} Q(p)^{\sss (i)}_{\alpha\beta\mu\nu}\,x_{i}^{\pi}\,
\log\left((-p^2-i0)/M^2\right)\Big\}\,;
\label{eq:b79}
\end{split}
\end{equation}
so that, with Eqs.~(\ref{eq:bc10}),~(\ref{eq:bc11}), we obtain
\begin{multline}
\hat{D}(p)^{tot}_{\alpha\beta\mu\nu}=\frac{1}{(2\pi)^2}\Big[+\frac{1}{2}\big( \eta_{\alpha\mu}\eta_{\beta\nu}+ \eta_{\alpha\nu}
\eta_{\beta\mu}\big)\,\frac{1}{-p^{2}-i0+f_{1}(p^2)}+\\-\frac{1}{2}\,\eta_{\alpha\beta}\eta_{\mu\nu}\,
\frac{1}{-p^{2}-i0+f_{2}(p^2)}\Big]+\text{non-contributing terms}\,.
\label{eq:b80}
\end{multline}
The non-contributing terms vanish between conserved energy-momentum matter tensors or between physical graviton
states~\cite{gri3}.

Mass normalization (the graviton mass remains zero after the radiative correction due to the self-energy) requires
\begin{equation}
f_{j}(p^2){\Big|}_{p^2 =0} =0\,,\quad j=1,2\,,
\label{eq:b81}
\end{equation}
and coupling constant normalization (the coupling constant is not changed by the radiative correction) requires
\begin{equation}
\frac{f_{j}(p^2)}{p^2}{\bigg|}_{p^2 =0} =0\,,\quad j=1,2\,.
\label{eq:b82}
\end{equation}
We work out these two conditions for the case $j=1$. Since $f_{1}(p^2)$ has the form (see Eqs.~(\ref{eq:b72}),~(\ref{eq:b78}),
~(\ref{eq:b79}),~(\ref{eq:bc8}) and~(\ref{eq:bc11}))
\begin{equation}
f_{1}(p^2)=-2\,(2\pi)^2\,\Xi\,\Big(-162\,p^{4}\,\log\big((-p^2-i0)/M^2\big)+c_{5}\,p^{2}+c_{10}\,p^4\Big)\,,
\label{eq:b83}
\end{equation}
the two conditions~(\ref{eq:b81}) and~(\ref{eq:b82}) are satisfied, if we set $c_{5}=0$. The analysis for the case
$j=2$ is much more involved due to the complexity of  Eqs.~(\ref{eq:bc8}) and~(\ref{eq:bc11}), although not conceptually difficult, 
and it yields the condition $c_{6}=0$. The remaining normalization constants are not determined by Eqs.~(\ref{eq:b81})
and~(\ref{eq:b82}), so that we are left with
\begin{equation}
\hat{\Pi}(p)^{\sss N}_{\alpha\beta\mu\nu}=\Xi\big[-656,-208,+162,-162,+118\big]\log\bigg(\frac{-p^2-i0}{M^2}\bigg)+
\hat{N}^{\sss (4)}(p)_{\alpha\beta\mu\nu},
\label{eq:b84}
\end{equation}
where $\hat{N}^{\sss (4)}(p)_{\alpha\beta\mu\nu}$ is a general SWI-invariant polynomial given in Eq.~(\ref{eq:b72})
with three free normalization constants. In order to simplify the above expression and to reduce  further  the
freedom in the  normalization, we rescale the constants $c_{7},c_{10}$ and $c_{11}$ in the following way
\begin{equation}
c_{11}=+118\log(M^2/K^2),\quad c_{7}=-656\log(M^2/L^2),\quad c_{10}=-162\log(M^2/N^2)\,;
\label{eq:b85}
\end{equation}
so that, with $(H^2)^{104}:=(N^2)^{81} (L^2)^{82}/(K^2)^{59}$, we obtain 
\begin{multline}
\hat{\Pi}(p)^{\sss N}_{\alpha\beta\mu\nu}=\Xi\,\bigg[-656\,\log\Big(\frac{-p^2-i0}{L^2}\Big),
-208\,\log\Big(\frac{-p^2-i0}{H^2}\Big),\\
+162\,\log\Big(\frac{-p^2-i0}{N^2}\Big),-162\,\log\Big(\frac{-p^2-i0}{N^2}\Big),
+118\,\log\Big(\frac{-p^2-i0}{K^2}\Big)\bigg]\,,
\label{eq:b86}
\end{multline}
with the four arbitrary positive masses $L,H,N$ and $K$. Now, since the splitting of the mass zero distribution 
$\hat{d}(p)=\Theta(p^2)\mathrm{sgn}(p^0)$, requires the introduction of a scale invariance breaking mass, 
it is natural to assume this mass scale to be unique, say $M_{0}$, which may correspond in case to the Planck mass, so 
that the graviton self-energy tensor including its  normalization now reads
\begin{equation}
\hat{\Pi}(p)^{\sss N}_{\alpha\beta\mu\nu}=\Xi\,\bigg[-656,-208,+162,-162,+118\bigg]\,
\log\bigg(\frac{-p^2-i0}{M_{0}^2}\bigg)\,.
\label{eq:b87}
\end{equation}
Therefore, the whole ambiguity in the normalization of $T_{2}(x,y)^{h\sss SE}$  may be reduced to the single parameter $M_{0}$ 
which remains present in the theory.

\section{Corrections to the Newtonian Potential}\label{sec:potential}
\setcounter{equation}{0}

In the last years a group of papers appeared~\cite{dono2},~\cite{hambe} and~\cite{akhu}, 
based on the proposal of treating perturbative quantum gravity as a low energy effective quantum theory. In this 
approach, leading quantum corrections to the Newtonian potential were reliably calculated in the long range, low energy 
limit. 

In this section we will show that causal perturbation theory for quantum gravity yields the same result for the leading
corrections to the Newtonian potential for heavy spinless particles described by the scalar field $\va$.

The Newtonian potential between two masses $m_{1},m_{2}$
\begin{equation}
V(r)=-\,G\,\frac{m_{1}m_{2}}{r}
\label{eq:b89}
\end{equation}
can be obtained in the static non-relativistic limit of a single graviton exchange tree diagram~\cite{dono2},~\cite{boul}
to lowest order in $G$, calculated from
\begin{equation}
T_{1}^{\sss M}(x)=i\,:\!\mathcal{L}^{\sss (1)}_{\sss M}(x)\!:=
i\,\frac{\kappa}{2}\,:\!h^{\alpha\beta}(x)\,
b_{\alpha\beta\mu\nu}\,T^{\mu\nu}_{m}(x)\!:\,,
\label{eq:b90}
\end{equation}
where the conserved energy-momentum matter tensor reads
\begin{equation}
T^{\mu\nu}_{m}(x)=\va(x)^{,\mu}\va(x)^{,\nu}-\eta^{\mu\nu}\,\mathcal{L}^{\sss (0)}_{\sss M}(x)\,,
\label{eq:b91}
\end{equation}
and the expansion in powers of the coupling constant of the matter Lagrangian density has the form
\begin{equation}
\mathcal{L}_{\sss M}=\frac{1}{2}\sqrt{-g}\,\big( g^{\mu\nu}\va_{;\mu}\va_{;\nu}-m^2\,\va^2\big)=\sum_{i=0}^{\infty}
\kappa^{i}\,\mathcal{L}^{\sss (i)}_{\sss M}\,.
\label{eq:b92}
\end{equation}

From Eq.~(\ref{eq:b90}), we carry out  the inductive construction of the $2$-point distribution which describes the 
$\va\va\to\va\va$ scattering through one-graviton exchange and find
\begin{equation}
T_{2}(x,y){\Big|}_{\va\va\to\va\va}=-\frac{\kappa^2}{4}\,:\!T^{\gamma\delta}_{m}(x)\,b_{\gamma\delta\alpha\beta}\,
T^{\rho\sigma}_{m}(y)\,b_{\rho\sigma\mu\nu}\!:\,\Big[ -i\,b^{\alpha\beta\mu\nu}\,D_{0}^{\sss F}(x-y)\Big]\,.
\label{eq:b93}
\end{equation}
The static non-relativistic limit of the right side  of~(\ref{eq:b93}) yields~(\ref{eq:b89}) using 
$T^{\mu\nu}_{m}\to(2\pi)\delta^{\mu 0}\delta^{\nu 0} m $ and the fact that the Fourier transform of the momentum 
transfer $\bol{p}^2$ reads $(4\pi r)^{-1}$. For a deeper understanding of the connection between $S$-matrix and
potential, see~\cite{sha}. Inserting one  self-energy contribution~(\ref{eq:b87}) on the graviton contraction of 
Eq.~(\ref{eq:b93}), we obtain
the order $\kappa^4$ correction
\begin{equation}
T_{2}(x,y)^{\text{corr}}=i\,\frac{\kappa^2}{4}\,:\!T^{\gamma\delta}_{m_{1}}(x)\,T^{\rho\sigma}_{m_{2}}(y)\!:\,
\Omega(x-y)^{\text{corr}}_{\gamma\delta\rho\sigma}\,,
\label{eq:b94}
\end{equation}
where, with the self-energy tensor of~(\ref{eq:b87}), we have
\begin{equation}
\hat{\Omega}(p)^{\text{corr}}_{\gamma\delta\rho\sigma}=2\frac{\hat{\Pi}(p)^{\sss N}_{\gamma\delta\rho\sigma}}{p^4}\,.
\label{eq:b95}
\end{equation}
Only the two last terms in the self-energy tensor will contribute in the following, because the others vanish
when paired with conserved matter energy-momentum tensors. Therefore, even if we had not assumed a unique mass scale
(as done at the end of Sec~\ref{sec:aufsumm}), the parameters $L$, $H$ and $N$ in~(\ref{eq:b86}) would not be 
physically observable. From~(\ref{eq:b93}) and~(\ref{eq:b95}) we obtain in the 
static non-relativistic limit the genuine quantum correction to the Newtonian potential
(inserting back the physical constants $\hbar$ and $c$)
\begin{equation}
V(r)=-\,G\,\frac{m_{1}m_{2}}{r}\,\left( 1+\frac{G\hbar}{c^{3}\pi}\,\frac{206}{30}\frac{1}{r^{2}}\right)\,.
\label{eq:b96}
\end{equation}
The central  piece in the calculation is the distributional Fourier transform of $\log\left(\bol{p}^{2}/M_{0}^{2}\right)$
which yields $(-2\pi r^3)^{-1}$ and the $M_{0}$-dependence disappears from the non-local part of the final result,
being proportional to $\delta^{\sss (3)}(\bol{x})$. The relevant length scale appearing in~(\ref{eq:b96}) is the Planck 
length $l_{p}=\sqrt{G\hbar/c^{3}}$.

Our result agrees with the corresponding one in~\cite{hambe}, although this represents only a partial correction to the 
Newtonian potential, because we have taken into account only the graviton self-energy contribution and not the complete 
set of  diagrams of order $\kappa^4$  contributing to these corrections, as, for example, the vertex correction or the double 
scattering. Therefore we cannot make any statement on the absolute sign of the correction in Eq.~(\ref{eq:b96}).
See also~\cite{od} for  the corrections to the Newtonian potential in the framework of $R^2$-theories.

A remark about different choices of $T_{1}^{h}$ or $T_{1}^{u}$: the graviton self-energy tensor of Eq.~(\ref{eq:b67}) leads 
to the same corrections, whereas that of Eq.~(\ref{eq:b68}), taking also the ghost loop contribution into account, leads to 
different corrections.

\section{Perturbative Gauge Invariance to Second Order for Loop Contributions}\label{sec:gauge}
\setcounter{equation}{0}

For the sake of completeness, before proving perturbative gauge invariance to second order in the loop graph sector,
we calculate the $2$-point ghost self-energy contributions.

\subsection{Ghost Self-Energy}\label{sec:ghose}

We follow the inductive causal construction described in Sec.~\ref{sec:twopoint} in order to construct the ghost 
self-energy contribution in second order perturbation theory. Starting with the ghost coupling $T_{1}^{u}(x)$ given in 
Eq.~(\ref{eq:b14}), we perform one graviton and one ghost--anti-ghost contraction, Eq.~(\ref{eq:b18}) in order to obtain 
the corresponding $D_{2}(x,y)$ distribution. The distribution splitting solution is the same as in 
Sec.~\ref{sec:splitting}, namely Eq.~(\ref{eq:b54}) with the singular order given by Eq.~(\ref{eq:b39}) in 
Sec.~\ref{sec:singord}. Therefore, after all these steps, we obtain
\begin{equation}
\begin{split}
T_{2}(x,y)=&+:\!u^{\gamma}(x)\tilde{u}^{\alpha}(y)_{,\beta}\!:\,t_{2}^{u,\sss(3a)}(x-y)_{\gamma\alpha}^{\quad\beta}+\\
              & +:\!\tilde{u}^{\alpha}(x)_{,\beta}u^{\gamma}(y)\!:\,t_{2}^{u,\sss(3b)}(x-y)_{\alpha\ \ \gamma}^{\ \ \beta}+\\
&+        :\!u^{\gamma}(x)_{,\beta}\tilde{u}^{\alpha}(y)_{,\nu}\!:\,t_{2}^{u,\sss(2a)}(x-y)_{\gamma\ \ \alpha}^{\ \beta\ \ \nu}+\\
& +       :\!\tilde{u}^{\alpha}(x)_{,\nu}u^{\gamma}(y)_{,\beta}\!:\,t_{2}^{u,\sss(2b)}(x-y)_{\alpha\ \ \gamma}^{\ \nu\ \  \beta}\,.
\label{eq:b97}
\end{split}
\end{equation}
The numerical distributions appearing here have already been introduced in Eq.~(\ref{eq:b17.1}) with the
notations: $t_{2}^{\sss u,(3a)}=i t_{\sss u \d \tilde{u}}$,
$t_{2}^{\sss u,(3b)}=i t_{\sss \d \tilde{u} u}$, $t_{2}^{\sss u,(2a)}=i t_{\sss \d u \d \tilde{u}}$ and
$t_{2}^{\sss u,(2b)}=i t_{\sss \d \tilde{u} \d u}$. 
The results for these ghost self-energy distributions are:
\begin{equation}
\begin{split}
\hat{t}_{2}^{\sss u,(3a)}(p)_{\gamma|\alpha\beta}=\hat{t}_{2}^{\sss u,(3b)}(p)_{\alpha\beta|\gamma}
=\Xi\,\big[&-80\,p_{\alpha}p_{\beta}p_{\gamma} +60\,p^2 p_{\gamma}\eta_{\alpha\beta} 
  -20\,p^2 p_{\alpha}\eta_{\beta\gamma}+\\ & -20\,p^2 p_{\beta}\eta_{\alpha\gamma}\big]\,
   \log\big( (-p^2 -i0)/ M^2 \big)\,,
\raisetag{5mm}\label{eq:b97.1}
\end{split}
\end{equation}
and
\begin{multline}
\hat{t}_{2}^{\sss u,(2a)}(p)_{\gamma\beta|\alpha\nu}=-\hat{t}_{2}^{\sss u,(2b)}(p)_{\alpha\nu|\gamma\beta}=
i\,\Xi\,\big[ +160 \,p^2 \eta_{\gamma\beta}\eta_{\alpha\nu} +240\,p_{\alpha} p_{\gamma}\eta_{\nu\beta}+\\
+240\, p_{\nu} p_{\gamma} \eta_{\alpha\beta} -240\, p_{\alpha}p_{\nu}\eta_{\gamma\beta}
-400\,p_{\gamma}p_{\beta}\eta_{\alpha\nu}\big]\,\log\big( (-p^2 -i0)/ M^2 \big)\,.
\label{eq:b97.2}
\end{multline}
By disregarding divergence terms after partial integration, we can recast Eq. (\ref{eq:b97}) into the form
\begin{equation}
T_{2}(x,y)^{g\sss SE}=:\!u^{\gamma}(x)\tilde{u}^{\alpha}(y)\!:\,i\,\Pi_{a}(x-y)_{\gamma\alpha}+
:\!\tilde{u}^{\alpha}(x)u^{\gamma}(y)\!:\,i\,\Pi_{b}(x-y)_{\alpha\gamma}\,.
\label{eq:b98}
\end{equation}
with the ghost self-energy tensors
\begin{equation}
\begin{split}
i\,\Pi_{a}(x-y)_{\gamma\alpha}:=&+\d_{\beta}^{x}t_{2}^{u,\sss(3a)}(x-y)_{\gamma\alpha}^{\quad\beta}
     -\d_{\beta}^{x}\d_{\nu}^{x}t_{2}^{u,\sss(2a)}(x-y)_{\gamma\ \ \alpha}^{\ \beta\ \ \nu}\,,\\
i\,\Pi_{b}(x-y)_{\alpha\gamma}:=&-\d_{\beta}^{x}t_{2}^{u,\sss(3b)}(x-y)_{\alpha\ \ \gamma}^{\ \ \beta}
      -\d_{\beta}^{x}\d_{\nu}^{x}t_{2}^{u,\sss(2b)}(x-y)_{\alpha\ \ \gamma}^{\ \nu\ \ \beta}\,;
\label{eq:b99}
\end{split}
\end{equation}
that  read in momentum space
\begin{equation}
\hat{\Pi}_{a}(p)_{\gamma\alpha}=\Xi\,\Big[40\,p^2 p_{\alpha}p_{\gamma}+20\,p^4 \eta_{\alpha\gamma}\Big]\,
                        \log\left(\frac{-p^2 -i0}{M^2}\right)=-\hat{\Pi}_{b}(p)_{\alpha\gamma}\,.
\label{eq:b100}
\end{equation}
We do not discuss here in detail the normalization  of this $2$-point distribution. But an investigation analogous 
to that of  Sec.~\ref{sec:aufsumm} for the sum of the series with an increasing number of  ghost self-energy insertions 
let us assume that the normalization terms with singular order smaller than four are set equal to zero, whereas those with 
singular order four may be absorbed in the scale invariance breaking mass $M$.

\subsection{Perturbative Gauge Invariance to Second Order for Loop Graphs}\label{sec:loop2}

The introduction of the $Q$-vertex $T_{1/1}^{\mu}(x)$, Eq.~(\ref{eq:b16}), enables us to formulate perturbative gauge invariance 
by means of Eq.~(\ref{eq:b17}). We call a theory gauge invariant to second order, if 
$T_{2}(x,y)$ satisfies
\begin{equation}
d_{Q} T_{2}(x,y)=\d_{\mu}^{x} T_{2/1}^{\mu}(x,y)+\d_{\mu}^{y} T_{2/2}^{\mu}(x,y)\,,
\label{eq:b106}
\end{equation}
where $T_{2/1}^{\mu}(x,y)\: \big(T_{2/2}^{\mu}(x,y)\big)$ is the two-point `renormalized' time-ordered  pro\-duct obtained 
by means of the inductive causal scheme with a $Q$-vertex at $x\, (y)$ and a normal vertex at $y\, (x)$. 

Since $R'_{2}(x,y)$ is trivially gauge invariant due to~(\ref{eq:b19}) and~(\ref{eq:b16}), it suffices to 
prove~(\ref{eq:b106}) with the retarded parts $R_{2}$, $R^{\mu}_{2/1}$ and $R^{\mu}_{2/2}$, instead of the corresponding 
$T$-distributions.

Taking Eq.~(\ref{eq:b20}) and~(\ref{eq:b16}) into account, we find that the $D_{2}$-distribution is 
trivially invariant
\begin{equation}
d_{Q} D_{2}(x,y)=\d_{\mu}^{x} D_{2/1}^{\mu}(x,y)+\d_{\mu}^{y} D_{2/2}^{\mu}(x,y)\,,
\label{eq:b107}
\end{equation}
with the definitions
\begin{equation}
\begin{split}
D_{2/1}^{\mu}(x,y)&:=\big[T_{1/1}^{\mu}(x),T_{1}(y)\big]\,,\\
D_{2/2}^{\mu}(x,y)&:=\big[T_{1}(x),T^{\mu}_{1/1}(y)\big]\,.
\label{eq:b108}
\end{split}
\end{equation}

The question is whether an equation similar to~(\ref{eq:b107}) holds for the retarded parts $R_{2}$, $R^{\mu}_{2/1}$ 
and $R^{\mu}_{2/2}$ of $D_{2}$, $D^{\mu}_{2/1}$  and $D^{\mu}_{2/2}$, respectively. In the inductive causal construction, 
the splitting of distributions in $D_{2}$, $D^{\mu}_{2/1}$ and $D^{\mu}_{2/2}$ may give rise to local normalization terms,
if the singular order is positive. We consider here only loop graphs in second order.  Therefore, we must show that
\begin{equation}
\begin{split}
d_{Q} R_{2}(x,y)^{\text{loops}}+d_{Q} N_{2}(x,y)^{\text{loops}}=&+\d_{\mu}^{x} R_{2/1}^{\mu}(x,y)^{\text{loops}}+
\d_{\mu}^{y} R_{2/2}^{\mu}(x,y)^{\text{loops}}+\\
&+\d_{\mu}^{x} N_{2/1}^{\mu}(x,y)^{\text{loops}}+\d_{\mu}^{y} N_{2/2}^{\mu}(x,y)^{\text{loops}}
\label{eq:b109}
\end{split}
\end{equation}
can be satisfied by a suitable choice of the free constants in the normalization terms $N_{2}^{\text{loops}}$, 
$N_{2/1}^{\mu\,\text{loops}}$ and $N_{2/2}^{\mu\,\text{loops}}$ of the general splitting solution of 
$D_{2}^{\text{loops}}$, $D^{\mu\,\text{loops}}_{2/1}$  and $D^{\mu\,\text{loops}}_{2/2}$, respectively.

Since every splitting solution agrees with the original distribution on the forward light cone $\overline{V^{+}}\backslash\{0\}$,
gauge invariance~(\ref{eq:b109})  can be violated by local terms with support $x=y$ only. Hence, the crucial point is 
the correct treatment of the local terms appearing in~(\ref{eq:b109}). 

A careful analysis shows that that the only local terms appearing in~(\ref{eq:b109}) are those belonging to the  
normalizations  $N_{2}^{\text{loops}}$, $N_{2/1}^{\mu\,\text{loops}}$ and $N_{2/2}^{\mu\,\text{loops}}$ of the 
distribution splitting. 

Let us analyze the different terms in~(\ref{eq:b109}). First of all, the normalization terms 
$N_{2}(x,y)^{\text{loops}}$ were already discussed in Sec.~\ref{sec:n2} and in Sec~\ref{sec:ghose}, below 
Eq.~(\ref{eq:b100}). They can be consistently set equal to zero.

The gauge variation of $R_{2}(x,y)^{\text{loops}}$ generates no local terms: the field operators in the normal products
undergo the infinitesimal gauge variations~(\ref{eq:b11.1}) and~(\ref{eq:b11.2}) and the numerical distributions 
remain unchanged. Taking an example from Eq.~(\ref{eq:b55})
\begin{equation}
\begin{split}
d_{Q}R_{2}(x,y)^{\text{loops}}&=d_{Q}\Big(:\!h^{\alpha\beta}(x)h^{\mu\nu}(y)\!:\,r_{2}^{\sss (4)}(x-y)_{\alpha\beta\mu\nu}
+\ldots\Big)\\&=:\!u^{\rho}(x)_{,\sigma}h^{\mu\nu}(y)\!:\,\big[-i\,b^{\alpha\beta\rho\sigma}\,
r_{2}^{\sss (4)}(x-y)_{\alpha\beta\mu\nu}\big]+\ldots\,.
\label{eq:b110}
\end{split}
\end{equation}
$r_{2}^{\sss (4)}(x-y)_{\alpha\beta\mu\nu}$ does not contain local terms, the same holds for $b^{\alpha\beta\rho\sigma}
r_{2}^{\sss (4)}(x-y)_{\alpha\beta\mu\nu}$.

Now we investigate the $R_{2/1}^{\mu}(x,y)^{\text{loops}}$-terms. Due to the great calculation complexity, we work out
only a representative example, that still contains the main features. Let us choose in $T_{1/1}^{\mu}(x)\sim uhh+\tilde{u}uu$ 
(from~\cite{scho1}) the term $-\frac{\kappa}{4}:\!u^{\gamma}(x)_{,\gamma}h(x)h(x)_{,\mu}\!:$ and in 
$T_{1}^{h+u}(y)$,~(\ref{eq:b6}), the term $-i\frac{\kappa}{4}:\!h^{\alpha\beta}(y)h(y)_{,\alpha}h(y)_{,\beta}\!:$.
We construct $D_{2/1}^{\mu}(x,y)^{\text{loops}}$ by carrying out two graviton contractions
\begin{multline}
D_{2/1}^{\mu}(x,y)^{\text{loops}}=i\,\frac{\kappa^2}{16}\bigg( +:\!u^{\gamma}(x)_{,\gamma}h^{\alpha\beta}(y)\!:
\big(-32\big[+D^{\sss (+)}_{\alpha|\mu\beta}-D^{\sss (-)}_{\alpha|\mu\beta}\big](x-y)\big)+\\
+:\!u^{\gamma}(x)_{,\gamma}h(y)_{,\alpha}\!:\big(+8\big[+D_{\cdot|\mu\alpha}^{\sss (+)}-D_{\cdot|\mu\alpha}^{\sss (-)}+
D_{\alpha|\mu}^{\sss (+)}-D_{\alpha|\mu}^{\sss (-)}\big](x-y)\big)\bigg)+\ldots\,,
\label{eq:b111}
\end{multline}
which can be written in the form
\begin{equation}
\begin{split}
D_{2/1}^{\mu}(x,y)^{\text{loops}}=&+:\!u^{\gamma}(x)_{,\gamma}h^{\alpha\beta}(y)\!:\,d_{2/1}^{\mu}(x-y)_{\alpha\beta}+\\
&+:\!u^{\gamma}(x)_{,\gamma}h(y)_{,\alpha}\!:\,d_{2/1}^{\mu}(x-y)^{\alpha}+\ldots\,,
\label{eq:b112}
\end{split}
\end{equation}
with the $\hat{D}_{\dots|\dots}^{\sss (\pm)}(p)$-functions given in App.~1:
\begin{gather}
\hat{d}_{2/1}^{\mu}(p)_{\alpha\beta}=-2i\,\kappa^2\,\big[+\hat{D}^{\sss (+)}_{\alpha|\mu\beta}-
       \hat{D}^{\sss (-)}_{\alpha|\mu\beta}\big](p)\,,\nonumber\\
\hat{d}_{2/1}^{\mu}(p)^{\alpha}=+i\,\frac{\kappa^2}{2}\,\big[+\hat{D}_{\cdot|\mu\alpha}^{\sss (+)}-
\hat{D}_{\cdot|\mu\alpha}^{\sss (-)}+
\hat{D}_{\alpha|\mu}^{\sss (+)}-\hat{D}_{\alpha|\mu}^{\sss (-)}\big](p)\,.
\label{eq:b113}
\end{gather}
Using the results of App.~1 and App.~2, we obtain
\begin{gather}
\hat{d}_{2/1}^{\mu}(p)_{\alpha\beta}=\frac{-\kappa^2\pi}{24(2\pi)^4}\big[2\,p_{\alpha}p_{\beta}p^{\mu}-
p^2 p_{\alpha}\eta_{\beta}^{\ \mu}+p^2 p_{\beta}\eta_{\alpha}^{\ \mu}+p^2 p^{\mu}\eta_{\alpha\beta}\big]\Theta(p^2)
\mathrm{sgn}(p^0),\nonumber\\
\hat{d}_{2/1}^{\mu}(p)^{\alpha}=\frac{+i\kappa^2\pi}{8(2\pi)^4}\,\big[p^{\mu}p^{\alpha}\big]\,\Theta(p^2)\,
\mathrm{sgn}(p^0)\,.
\label{eq:b114}
\end{gather}
In order to obtain $R_{2/1}^{\mu\,\text{loops}}$, we split the distributions in~(\ref{eq:b114}). With~(\ref{eq:b53})
and assuming that the splitting of the massless distribution $\hat{d}(p)$ generates only one mass scale $M$, we find
\begin{gather}
\hat{r}_{2/1}^{\mu}(p)_{\alpha\beta}=\frac{-i\kappa^2\pi}{24(2\pi)^5}\big[\text{the same as in~(\ref{eq:b114})}\big]
\log\left(\frac{-p^2-ip^0 0}{M^2}\right)\,,\nonumber\\
\hat{r}_{2/1}^{\mu}(p)^{\alpha}=\frac{-\kappa^2\pi}{8(2\pi)^5}\,\big[p^{\mu}p^{\alpha}\big]\,
\log\left(\frac{-p^2-ip^0 0}{M^2}\right)\,,
\label{eq:b115}
\end{gather}
so that 
\begin{equation}
\begin{split}
R_{2/1}^{\mu}(x,y)^{\text{loops}}=&+:\!u^{\gamma}(x)_{,\gamma}h^{\alpha\beta}(y)\!:\,r_{2/1}^{\mu}(x-y)_{\alpha\beta}+\\
&+:\!u^{\gamma}(x)_{,\gamma}h(y)_{,\alpha}\!:\,r_{2/1}^{\mu}(x-y)^{\alpha}+\ldots\,.
\label{eq:b116}
\end{split}
\end{equation}
The local normalization terms of the splitting are included in $N_{2/1}^{\mu}(x,y)^{\text{loops}}$ and reads:
\begin{equation}
\begin{split}
N_{2/1}^{\mu}(x,y)^{\text{loops}}=&+:\!u^{\gamma}(x)_{,\gamma}h^{\alpha\beta}(y)\!:\,n_{2/1}^{\mu}(x-y)_{\alpha\beta}+\\
&+:\!u^{\gamma}(x)_{,\gamma}h(y)_{,\alpha}\!:\,n_{2/1}^{\mu}(x-y)^{\alpha}+\ldots\,.
\label{eq:b117}
\end{split}
\end{equation}
Here, the $n_{2/1}$-distributions contain free normalization constants. The terms $R_{2/2}^{\mu}(x,y)^{\text{loops}}$
and $N_{2/2}^{\mu}(x,y)^{\text{loops}}$ coming from the second commutator in~(\ref{eq:b108}) can be analogously
calculated with $x\leftrightarrow y$ and an overall sign change.
Applying now $\d_{\mu}^{x}$ to $R_{2/1}^{\mu}(x,y)^{\text{loops}}$, we obtain
\begin{multline}
+:\!u^{\gamma}(x)_{,\gamma\mu}h^{\alpha\beta}(y)\!:\,r_{2/1}^{\mu}(x-y)_{\alpha\beta}
+:\!u^{\gamma}(x)_{,\gamma\mu}h(y)_{,\alpha}\!:\,r_{2/1}^{\mu}(x-y)^{\alpha}+\\
+:\!u^{\gamma}(x)_{,\gamma}h^{\alpha\beta}(y)\!:\,\d_{\mu}^{x}\,r_{2/1}^{\mu}(x-y)_{\alpha\beta}
+:\!u^{\gamma}(x)_{,\gamma}h(y)_{,\alpha}\!:\,\d_{\mu}^{x}\,r_{2/1}^{\mu}(x-y)^{\alpha}\,.
\label{eq:b118}
\end{multline}
The first two addends do not contain local terms, because the derivatives act only on the external fields and the
numerical distributions do not contain local terms by construction. The derivative acting on the numerical distributions
in the last two terms changes them into
\begin{gather}
-i\,p_{\mu}\, \hat{r}_{2/1}^{\mu}(p)_{\alpha\beta}=\frac{-\kappa^2\pi}{24(2\pi)^5}\big[+2p^2 p_{\alpha}p_{\beta}+
p^4\eta_{\alpha\beta}\big]\,\log\left(\frac{-p^2-ip^0 0}{M^2}\right)\,,\nonumber\\
-i\,p_{\mu}\, \hat{r}_{2/1}^{\mu}(p)^{\alpha}=\frac{i\kappa^2 \pi}{8(2\pi)^5}\,\big[p^2 p_{\alpha}\big]\,
\log\left(\frac{-p^2-ip^0 0}{M^2}\right)\,.
\label{eq:b119}
\end{gather}
Since these terms are proportional to the logarithms, none of them is local. The local anomaly producing mechanism
described in~\cite{scho1} does  not apply in the case of loop graphs. This is a consequence of the presence in loop 
graphs of products of Jordan--Pauli distributions that yield  non-local terms after distribution splitting.

We turn now to the local terms in $\d_{\mu}^{x}N^{\mu}_{2/1}(x,y)^{\text{loops}}$. Since 
$\omega(\hat{d}_{2/1}^{\mu\alpha\beta})=3$ and $\omega(\hat{d}_{2/1}^{\mu\alpha})=2$, these terms have the general form
\begin{equation}
\begin{split}
\hat{n}_{2/1}^{\mu}(p)^{\alpha\beta}&=+a_{1}p^{\mu}\eta^{\alpha\beta}+a_{2}p^{\alpha}\eta^{\mu\beta}
   +a_{3}p^{\beta}\eta^{\mu\alpha}+a_{4}p^{\mu}p^{\alpha}p^{\beta}+\\
&\quad +p^2\,\big(a_{5}p^{\mu}\eta^{\alpha\beta}
+a_{6}p^{\alpha}\eta^{\beta\mu}+a_{7}p^{\beta}\eta^{\alpha\mu}\big)\,,\\
\hat{n}_{2/1}^{\mu}(p)^{\alpha}&=+a_{8}\eta^{\mu\alpha}+a_{9}p^{\alpha}p^{\mu}+a_{10}p^2 \eta^{\alpha\mu}\,,
\label{eq:b120}
\end{split}
\end{equation}
where $a_{1},\ldots,a_{10}$ are unknown parameters. Multiplying~(\ref{eq:b120}) with $p_{\mu}$ leads to 
\begin{equation}
\begin{split}
p_{\mu}\,\hat{n}_{2/1}^{\mu}(p)^{\alpha\beta}&=+a_{1}p^2\eta^{\alpha\beta}+(a_{2}+a_{3})p^{\alpha}p^{\beta}
+(a_{4}+a_{6}+a_{7})p^2 p^{\alpha} p^{\beta}+a_{5} p^4 \eta^{\alpha\beta}\,,\\
p_{\mu}\,\hat{n}_{2/1}^{\mu\alpha}(p)&=+a_{8}p^{\alpha}+(a_{9}+a_{10})p^2 p^{\alpha}\,.
\label{eq:b121}
\end{split}
\end{equation}
Therefore, the  only local terms in~(\ref{eq:b109}) are those coming from~(\ref{eq:b121}). In our simplified example,
perturbative gauge invariance then requires
\begin{equation}
a_{1}=a_{2}+a_{3}=a_{4}+a_{6}+a_{7}=a_{5}=a_{8}=a_{9}+a_{10}=0\,,
\label{eq:b122}
\end{equation}
Obviously, these conditions are fulfilled by choosing $a_{1}=\ldots = a_{10}=0$. One may convince oneself that the above
example can be generalized to all second order loop graph contributions, because these follow the same pattern, although
much more involved conditions as~(\ref{eq:b122}) would appear. The important point is that
$\d_{\mu}^{x} R_{2/1}^{\mu}(x,y)^{\text{loops}}+\d_{\mu}^{y} R_{2/2}^{\mu}(x,y)^{\text{loops}}$ does not generate
local terms. This is in contrast to the tree graph sector, investigated in~\cite{scho1}. 

Let us add a final remark:
if we do not fix the normalization  $N_{2}(x,y)^{\text{loops}}$ of  $R_{2}(x,y)^{\text{loops}}$ (as done
in Sec.~\ref{sec:norm}), then the condition of perturbative gauge invariance to second order forces us to choose
the normalization constants $c_{i}$ of $N_{2}(x,y)^{\text{loops}}$ and $a_{j}$ of $N_{2/1+2}^{\mu}(x,y)^{\text{loops}}$
in such a way that 
\begin{equation}
d_{Q} N_{2}(x,y)^{\text{loops}}=+\d_{\mu}^{x} N_{2/1}^{\mu}(x,y)^{\text{loops}}+
\d_{\mu}^{y} N_{2/2}^{\mu}(x,y)^{\text{loops}}
\label{eq:b123}
\end{equation}
holds. Since~(\ref{eq:b109}) always holds among non-local terms by construction, a trivial solution $c_{i}=a_{j}=0$
$\forall i,j$ always exists. This concludes the proof of perturbative gauge invariance to second order for loop graphs.

\section*{Acknowledgements}

I would like to thank Prof.~G.~Scharf for his continuous and patient support, Adrian M\"uller and Mark
Wellmann  for stimulating discussions. 

\addcontentsline{toc}{section}{Appendix}

\addcontentsline{toc}{subsection}{Appendix 1: The \boldmath{$\hat{D}^{\sss(\pm)}_{\alpha\beta|\mu\nu}(p)$}-Functions}

\section*{Appendix 1: The \boldmath{$\hat{D}^{\sss(\pm)}_{\alpha\beta|\mu\nu}(p)$}-Functions}
\renewcommand{\theequation}{A.\arabic{equation}}
\setcounter{equation}{0}

The product of two $D_{0}^{\sss (\pm)}$-distributions is well-defined in momentum space, because the intersection of the 
supports of the two $\hat{D}_{0}^{\sss (\pm)}(p)$ is a compact set. The product 
$D^{\sss(\pm)}_{\cdot|\cdot}(x):=D_{0}^{\sss(\pm)}(x)\cdot D_{0}^{\sss(\pm)}(x)$ goes 
over into a convolution of the Fourier transforms $\hat{D}_{0}^{\sss (\pm)}(p)$ using Eq.~(\ref{eq:b9}):
\begin{equation}
\begin{split}
\hat{D}^{\sss(\pm)}_{\cdot|\cdot}(p)&=\frac{1}{(2\pi)^{2}}\int\!\! d^{4}x\,D_{0}^{\sss(\pm)}(x)\cdot 
D_{0}^{\sss(\pm)}(x)\,e^{i\,p\cdot x} \\
&=\frac{-1}{(2\pi)^8}\int\!\! d^{4}p_{1}\,d^{4}p_{2}\,\delta(p_{1}^{2})\,\Theta(\pm p_{1}^{0})\,
\delta(p_{2}^{2})\,\Theta(\pm p_{2}^{0})\cdot\int\!\!d^{4}x\,e^{-i\, x\cdot (p_{1}+p_{2}-p )}\\
&=\frac{-1}{(2\pi)^4}\int\!\! d^{4}k \,\delta\big( (p-k)^2 \big)\,\Theta\big(\pm (p^{0}-k^{0})\big)\,
\delta(k^{2})\,\Theta(\pm k^{0})\,.
\label{eq:ba1}
\end{split}
\end{equation}
If derivatives are acting on the $D_{0}^{\sss(\pm)}(x)$-distributions, then we define
\begin{gather}
D_{\alpha |\beta}^{\sss (\pm)}(x):=\partial_{\alpha}^{x} D_{0}^{\sss(\pm)}(x)\cdot\partial_{\beta}^{x} 
D_{0}^{\sss (\pm)}(x)\,,\nonumber\\
D_{\alpha\beta |\mu\nu}^{\sss (\pm)}(x):=\partial_{\alpha}^{x}\partial_{\beta}^{x} D_{0}^{\sss (\pm)}(x)
\cdot \partial_{\mu}^{x}\partial_{\nu}^{x} D_{0}^{\sss (\pm)}(x)\,,
\label{eq:ba2}
\end{gather}
and so on if a different combination of derivatives acts on the distributions. Following the same calculation as 
in~(\ref{eq:ba1}), 
we obtain in momentum space:
\begin{equation}
\begin{split}
\hat{D}_{\alpha |\beta}^{\sss (\pm)}(p)=\frac{+1}{(2\pi)^4}\int\!\! d^{4}k \,&\delta\big( (p-k)^2 \big)\,
\Theta\big(\pm (p^{0}-k^{0})\big)\,\delta(k^{2})\,\Theta(\pm k^{0})\\
&\times\Big[p_{\alpha}k_{\beta}-k_{\alpha}k_{\beta}\Big]\,,\nonumber 
\end{split}
\end{equation}
\begin{multline}
\hat{D}_{\alpha\beta |\mu\nu}^{\sss (\pm)}(p)=\frac{-1}{(2\pi)^4}\int\!\! d^{4}k \,\delta\big( (p-k)^2 \big)\,
\Theta\big(\pm (p^{0}-k^{0})\big)\,\delta(k^{2})\,\Theta(\pm k^{0}) \\
\times\Big[+p_{\alpha}p_{\beta}k_{\mu}k_{\nu}
-p_{\alpha}k_{\beta}k_{\mu}k_{\nu}-p_{\beta}k_{\alpha}k_{\mu}k_{\nu}+k_{\alpha}k_{\beta}k_{\mu}k_{\nu}\Big].
\raisetag{2cm}\label{eq:ba3}
\end{multline}
Therefore we see that we have to deal with integrals of the type
\begin{equation}
\begin{split}
I^{\sss (\pm)}(p)_{- / \alpha/\alpha\beta/\alpha\beta\mu/\alpha\beta\mu\nu}:=\int\!\! d^{4}k
\,&\delta\big( (p-k)^2 \big)\,
\Theta\big(\pm (p^{0}-k^{0})\big)\,\delta(k^{2})\,\Theta(\pm k^{0}) \\
&\times\big[1,k_{\alpha},k_{\alpha}k_{\beta},
k_{\alpha}k_{\beta}k_{\mu},k_{\alpha}k_{\beta}k_{\mu}k_{\nu}\big]\,,
\label{eq:ba4}
\end{split}
\end{equation}
which are calculated in App.~2. For the two examples in Eq.~(\ref{eq:ba3}), we find the relations
\begin{gather}
\hat{D}_{\cdot|\cdot}^{\sss(\pm)}(p)=\frac{-1}{(2\pi)^{4}}\,I^{\sss (\pm)}(p)\,, \nonumber\\
\hat{D}_{\alpha |\beta}^{\sss (\pm)}(p)=\frac{+1}{(2\pi)^4}\Big[+p_{\alpha}\,I^{\sss(\pm)}(p)_{\beta}-
I^{\sss(\pm)}(p)_{\alpha\beta}\Big]\,,\nonumber\\
\begin{split}
\hat{D}_{\alpha\beta |\mu\nu}^{\sss(\pm)}(p)=\frac{-1}{(2\pi)^4}\Big[&+p_{\alpha}p_{\beta}\,I^{\sss(\pm)}(p)_{\mu\nu}
-p_{\alpha}\,I^{\sss(\pm)}(p)_{\beta\mu\nu}\\
&-p_{\beta}\,I^{\sss(\pm)}(p)_{\alpha\mu\nu}+I^{\sss(\pm)}(p)_{\alpha\beta\mu\nu}\Big]\,,
\label{eq:ba5}
\end{split}
\end{gather}
between the $\hat{D}^{\sss(\pm)}_{\ldots|\ldots}(p)$-functions and the $I^{\sss(\pm)}(p)_{\ldots}$-integrals.

\addcontentsline{toc}{subsection}{Appendix 2: The \boldmath{$I^{\sss(\pm)}(p)_{\ldots}$}-Integrals}

\section*{Appendix 2: The \boldmath{$I^{\sss(\pm)}(p)_{\ldots}$}-Integrals}
\renewcommand{\theequation}{B.\arabic{equation}}
\setcounter{equation}{0}

For the case of $I^{\sss(+)}(p)$, Eq.~(\ref{eq:ba4}), the momenta $k$ and $p$ are restricted to the space-time regions
$\{k^2 =0, k_0 > 0\}$ and $\{(p-k)^2 =0, p_0-k_0 > 0\}$, due to the $\delta$- and $\Theta$-distributions in the integrand. 
Then $p-k$ and $k$ are time-like and therefore $p$ is  time-like. We 
choose a Lorentz reference frame with $p_{\alpha}=(p_0 , \bol{0}),\,p^{0}>0$, then
\begin{equation}
I^{\sss (+)}(p_0 )=\int\!\! d^{4}k\, \delta(p_{0}^{2}-2p_{0}k_{0})\, \Theta(p_{0}-k_{0})\,\Theta(k_{0})\,
\frac{\delta(k_{0}-|\bol{k}|)+\delta(k_{0}+|\bol{k}|)}{2\,E_{\bol{k}}}
\label{eq:bb1}
\end{equation}
with $E_{\bol{k}}=k_{0}=|\bol{k}|$, so that
\begin{equation}
\begin{split}
I^{\sss(+)}(p_{0})& =4\pi\int_{-\infty}^{+\infty}\!\!dk^{0}\int_{0}^{+\infty}\!\!d|\bol{k}|\,|\bol{k}|^2 \, 
\delta\big(2p_{0} (\frac{p_{0}}{2}-k_{0})\big)\,\Theta(p_{0}-k_{0})\,\frac{\delta(k_{0}-|\bol{k}|)}{2|\bol{k}|} \\
       & =\frac{\pi}{p_{0}}\int_{-\infty}^{+\infty}\!\!dk^{0}\int_{0}^{+\infty}\!\!d|\bol{k}|\,|\bol{k}|\,
 \delta(\frac{p_{0}}{2}-k_{0})\,\Theta(p_{0}-k_{0})\,\delta(k_{0}-|\bol{k}|)\\
& =\frac{\pi}{p_{0}}\int_{0}^{p_{0}}\!\!d|\bol{k}|\,|\bol{k}|\,\delta(|\bol{k}|-\frac{p_{0}}{2})=\frac{\pi}{2}\,.
\label{eq:bb2}
\end{split}
\end{equation}
$I^{\sss(-)}(p)$ can be calculated analogously and the result  in an arbitrary Lorentz reference frame is
\begin{equation}
I^{\sss(\pm)}(p)=\frac{\pi}{2}\,\Theta(p^2)\,\Theta(\pm p^{0})\,.
\label{eq:bb3}
\end{equation}

Computing  $I^{\sss(\pm)}(p)_{\alpha}$ for  $p_{\alpha}=(p_0 , \bol{0}),\,p_{0}>0$, we have a non-vanishing contribution only 
for $\alpha=0$. We obtain for $I^{\sss(\pm)}(p_0)_{0}$ an additional factor $k_{0}$ in the integrand of~(\ref{eq:bb2}), 
which is set equal to $|\bol{k}|$, because of the distribution $\delta(k_{0}-|\bol{k}|)$ and finally is set equal to 
$p_{0}/2$, because of the distribution $\delta(|\bol{k}|-\frac{p_{0}}{2})$. This leads to
\begin{equation}
I^{\sss(\pm)}(p)_{\alpha}=\frac{\pi}{4}\,p_{\alpha}\,\Theta(p^2)\,\Theta(\pm p^{0})\,,
\label{eq:bb4}
\end{equation}
in an arbitrary Lorentz reference frame. 

For $I^{\sss(\pm)}(p)_{\alpha\beta}$, we consider the covariant decomposition
\begin{equation}
I^{\sss(\pm)}(p)_{\alpha\beta}=A^{\sss(\pm)}(p^2)\,p_{\alpha}p_{\beta}+B^{\sss(\pm)}(p^2)\,\eta_{\alpha\beta}\,.
\label{eq:bb5}
\end{equation}
It follows from $I^{\sss(\pm)}(p)_{\alpha}^{\ \alpha}=0$ (because of the factor $k^2 \delta(k^2)$ in the integrand of 
Eq.~(\ref{eq:ba4})), that
\begin{equation}
B^{\sss(\pm)}(p^2)=\frac{-p^2}{4}\,A^{\sss(\pm)}(p^2)\,.
\label{eq:bb6}
\end{equation}
Then
\begin{equation}
I^{\sss(\pm)}(p)_{\alpha\beta}\,p^{\alpha}p^{\beta}=\frac{3}{4}\,A^{\sss(\pm)}(p^2)\,p^4\,.
\label{eq:bb7}
\end{equation}
Calculating $I^{\sss(+)}(p)_{\alpha\beta}\,p^{\alpha}p^{\beta}$ for $p_{\alpha}=(p_{0},\bol{0}),\,p_{0}>0$, through 
the integral definition~(\ref{eq:ba4}), an additional factor $(p_{0}k_{0})^2$ appears in the integrand, therefore we 
obtain in an arbitrary Lorentz frame
\begin{equation}
I^{\sss(\pm)}(p)_{\alpha\beta}\,p^{\alpha}p^{\beta}=\frac{\pi}{8}\,p^4\,\Theta(p^2)\,\Theta(\pm p^{0})\,.
\label{eq:bb8}
\end{equation}
Comparing~(\ref{eq:bb7}) with~(\ref{eq:bb8}) we find $A^{\sss(\pm)}(p^2)=\frac{\pi}{6}\,\Theta(p^2)\,\Theta(\pm p^{0})$, 
so that
\begin{equation}
I^{\sss(\pm)}(p)_{\alpha\beta}=\frac{\pi}{6}\,\Big(p_{\alpha}p_{\beta}-\frac{p^2}{4}\,\eta_{\alpha\beta}\Big)
\,\Theta(p^2)\,\Theta(\pm p^{0})\,.
\label{eq:bb9}
\end{equation}

For $I^{\sss(\pm)}(p)_{\alpha\beta\mu}$, if we calculate $I^{\sss(+)}(p)_{\alpha\beta\mu}p^{\alpha}p^{\beta}p^{\mu}$ for
$p_{\alpha}=(p_0 , \bol{0}),\,p_{0}>0$, we get a factor $(p_{0}k_{0})^3$ in the integrand, so that in an arbitrary 
Lorentz frame we have
\begin{equation}
I^{\sss(\pm)}(p)_{\alpha\beta\mu}\,p^{\alpha}p^{\beta}p^{\mu}=\frac{\pi}{16}\,p^6\,\Theta(p^2)\,\Theta(\pm p^{0})\,.
\label{eq:bb10}
\end{equation}
The covariant decomposition of $I^{\sss(\pm)}(p)_{\alpha\beta\mu}$ reads
\begin{equation}
I^{\sss(\pm)}(p)_{\alpha\beta\mu}=C^{\sss(\pm)}(p^2)\,p^{\alpha}p^{\beta}p^{\mu}+D^{\sss(\pm)}(p^2)\,
\Big(p_{\alpha}\,\eta_{\beta\mu}+p_{\beta}\,\eta_{\alpha\mu}+p_{\mu}\,\eta_{\alpha\beta}\Big)\,.
\label{eq:bb11}
\end{equation}
Since $I^{\sss(\pm)}(p)_{\alpha\beta}^{\quad\beta}=0$, we obtain $D^{\sss(\pm)}(p^2)=\frac{-p^2}{6}\,C^{\sss(\pm)}(p^2)$. 
On the other side, contracting the covariant decomposition of $I^{\sss(\pm)}(p)_{\alpha\beta\mu}$ with 
$p^{\alpha}p^{\beta}p^{\mu}$, we find
\begin{equation}
I^{\sss(\pm)}(p)_{\alpha\beta\mu}\,p^{\alpha}p^{\beta}p^{\mu}=\frac{p^6}{2}\,C^{\sss(\pm)}(p^2)\,.
\label{eq:bb12}
\end{equation}
Comparing~(\ref{eq:bb12}) with~(\ref{eq:bb10}) we conclude that $C^{\sss(\pm)}(p^2)=\frac{\pi}{8}\Theta(p^2)\,
\Theta(\pm p^{0})$ and
\begin{equation}
I^{\sss(\pm)}(p)_{\alpha\beta\mu}=\frac{\pi}{8}\,\bigg(+ p_{\alpha}p_{\beta}p_{\mu}-\frac{p^2}{6}\Big(p_{\alpha}\,
\eta_{\beta\mu}+p_{\beta}\,\eta_{\alpha\mu}+p_{\mu}\,\eta_{\alpha\beta}\Big)\bigg)\,\Theta(p^2)\,\Theta(\pm p^{0})\,.
\label{eq:bb14}
\end{equation}

We repeat this calculation scheme also for $I^{\sss(\pm)}(p)_{\alpha\beta\mu\nu}$, which has the covariant decomposition
\begin{multline}
I^{\sss(\pm)}(p)_{\alpha\beta\mu\nu}=E^{\sss(\pm)}(p^2)\,p_{\alpha}p_{\beta}p_{\mu}p_{\nu}+F^{\sss(\pm)}(p^2)\,
\Big(+p_{\alpha}p_{\beta}\,\eta_{\mu\nu}+p_{\alpha}p_{\mu}\,\eta_{\beta\nu}+p_{\alpha}p_{\nu}\,\eta_{\beta\mu}+\\
     +p_{\beta}p_{\mu}\,\eta_{\alpha\nu}+p_{\beta}p_{\nu}\,\eta_{\alpha\mu}+p_{\mu}p_{\nu}\,\eta_{\alpha\beta}\Big)
+G^{\sss(\pm)}(p^2)\,\Big(+\eta_{\alpha\mu}\eta_{\beta\nu}+\eta_{\alpha\nu}\eta_{\beta\nu}+\eta_{\alpha\beta}\eta_{\mu\nu}
\Big)\,.
\label{eq:bb15}
\end{multline}
From $I^{\sss(\pm)}(p)^{\alpha}_{\ \alpha\mu\nu}=0$ and $I^{\sss(\pm)}(p)^{\alpha\quad\mu}_{\ \alpha\mu}=0$, we obtain
\begin{equation}
\begin{cases}
& F^{\sss(\pm)}(p^2)=\frac{-p^2}{8}\,E^{\sss(\pm)}(p^2)\,, \\
& G^{\sss(\pm)}(p^2)=\frac{-p^2}{6}\,F^{\sss(\pm)}(p^2)=\frac{p^4}{48}\,E^{\sss(\pm)}(p^2)\,.
\end{cases}
\label{eq:bb16}
\end{equation}
Computing $I^{\sss(+)}(p)_{\alpha\beta\mu\nu}\,p^{\alpha}p^{\beta}p^{\mu}p^{\nu}$, for 
$p_{\alpha}=(p_0 ,\bol{0}),\,p_{0}>0$, we get a factor $(p_{0}k_{0})^4$ in the integrand, so that in an arbitrary  
Lorentz frame we have
\begin{equation}
I^{\sss(\pm)}(p)_{\alpha\beta\mu\nu}\,p^{\alpha}p^{\beta}p^{\mu}p^{\nu}=\frac{\pi}{32}\,p^8\,
\Theta(p^2)\,\Theta(\pm p^{0})\,.
\label{eq:bb17}
\end{equation}
On the other side, contracting the covariant decomposition of $I^{\sss(\pm)}(p)_{\alpha\beta\mu\nu}$ with 
$p^{\alpha}p^{\beta}p^{\mu}p^{\nu}$ and using~(\ref{eq:bb16}), we obtain
\begin{equation}
I^{\sss(\pm)}(p)_{\alpha\beta\mu\nu}\,p^{\alpha}p^{\beta}p^{\mu}p^{\nu}=\frac{5}{16}\,p^8\,E^{\sss(\pm)}(p^2)\,.
\label{eq:bb18}
\end{equation}
Comparing~(\ref{eq:bb18}) with~(\ref{eq:bb17}), we find $E^{\sss(\pm)}(p^2)=\frac{\pi}{10}\Theta(p^2)\Theta(\pm p^{0})$, 
that implies with~(\ref{eq:bb16}) $F^{\sss(\pm)}(p^2)=\frac{-\pi p^2}{80}\Theta(p^2)\Theta(\pm p^{0})$ and 
$G^{\sss(\pm)}(p^2)=\frac{+\pi p^4}{480}\Theta(p^2)\Theta(\pm p^{0})$ so that
\begin{multline}
I^{\sss(\pm)}(p)_{\alpha\beta\mu\nu}=\frac{\pi}{10}\,\bigg(+p_{\alpha}p_{\beta}p_{\mu}p_{\nu}-\frac{p^2}{8}\,
\Big(+p_{\alpha}p_{\beta}\,\eta_{\mu\nu}+p_{\alpha}p_{\mu}\,\eta_{\beta\nu}+\\
+p_{\alpha}p_{\nu}\,\eta_{\beta\mu}+ p_{\beta}p_{\mu}\,\eta_{\alpha\nu}+p_{\beta}p_{\nu}\,\eta_{\alpha\mu}
+p_{\mu}p_{\nu}\,\eta_{\alpha\beta}\Big)+\\
+\frac{p^4}{48}\,\Big(+\eta_{\alpha\mu}\eta_{\beta\nu}+\eta_{\alpha\nu}\eta_{\beta\nu}+\eta_{\alpha\beta}\eta_{\mu\nu}
\Big)\bigg)\,\Theta(p^2)\,\Theta(\pm p^{0}) \,.
\label{eq:bb19}
\end{multline}

\addcontentsline{toc}{subsection}{Appendix 3: The \boldmath{$Q^{\sss (i)}(p)_{\alpha\beta,\mu\nu}$}-Projection Operators}

\section*{Appendix 3: The \boldmath{$Q^{\sss (i)}(p)_{\alpha\beta,\mu\nu}$}-Projection Operators}
\renewcommand{\theequation}{C.\arabic{equation}}
\setcounter{equation}{0}

The aim of this Appendix is to find a representation basis for rank-4 tensors, which allows to compute the inverse of the 
total graviton propagator~(\ref{eq:b77}) in Sec.~\ref{sec:norm}. Let us define as in~\cite{ste1}
\begin{equation}
d_{\mu\nu}:=\eta_{\mu\nu}-\frac{k_{\mu}k_{\nu}}{k^2}\,,\quad e_{\mu\nu}:=\frac{k_{\mu}k_{\nu}}{k^2}\,;
\label{eq:bc1}
\end{equation}
with 
\begin{equation}
d_{\mu\nu}d_{\nu\rho}=d_{\mu\rho}\,,\quad d_{\mu\nu}e_{\nu\rho}=0\,,\quad e_{\mu\nu}e_{\nu\rho}=e_{\mu\rho}\,.
\label{eq:bc2}
\end{equation}
Then the so-called `projection' operators are defined by
\begin{gather}
Q(k)^{\sss(1)}_{\alpha\beta,\mu\nu}=\frac{1}{2}\big( d_{\alpha\mu}e_{\beta\nu} +  d_{\alpha\nu}e_{\beta\mu} + 
        d_{\beta\mu}e_{\alpha\nu}+   d_{\beta\nu}e_{\alpha\mu}\big) \,,\nonumber\\
Q(k)^{\sss(2)}_{\alpha\beta,\mu\nu}=\frac{1}{2}\big(d_{\alpha\mu}d_{\beta\nu}+d_{\alpha\nu}d_{\beta\mu}-
       \frac{2}{3}d_{\alpha\beta}d_{\mu\nu}\big)\,,\nonumber\\
Q(k)^{\sss(3)}_{\alpha\beta,\mu\nu}=\frac{1}{3}\big(d_{\alpha\beta}d_{\mu\nu}\big)\,,\quad 
Q(k)^{\sss(4)}_{\alpha\beta,\mu\nu}=\big(e_{\alpha\beta}e_{\mu\nu}\big)\,;
\label{eq:bc3}
\end{gather}
and the so-called `transfer' operators are defined by
\begin{equation}
Q(k)^{\sss(5)}_{\alpha\beta,\mu\nu}=\frac{1}{\sqrt{3}}\big(d_{\alpha\beta}e_{\mu\nu}\big)\,,\quad 
Q(k)^{\sss(6)}_{\alpha\beta,\mu\nu}=\frac{1}{\sqrt{3}}\big(e_{\alpha\beta}d_{\mu\nu}\big)\,;
\label{eq:bc4}
\end{equation}
with the relations
\begin{equation}
Q(k)^{\sss(j)}_{\alpha\beta,\mu\nu}\,Q(k)^{\sss(j)}_{\mu\nu,\rho\sigma}=\begin{cases}
Q(k)^{\sss(j)}_{\alpha\beta,\rho\sigma}\,,& \text{if $j=1,2,3,4$}; \\
0\,, &  \text{if $j=5,6$}.
\end{cases}
\label{eq:bc5}
\end{equation}
These and other relations can be easily calculated using~(\ref{eq:bc2}). We consider a  rank-4 tensor in the standard 
basis as in Eq.~(\ref{eq:b59}) or~(\ref{eq:b60}) by giving its five coefficients and disregarding the logarithmic 
dependence on $k^2 / M^2$:
\begin{equation}
T(k)_{\alpha\beta\mu\nu}=\big[A,B,C,E,F\big](k)_{\alpha\beta\mu\nu}\,.
\label{eq:bc6}
\end{equation}
We rescale it by dividing it by $k^4$, so that we can express the obtained `rescaled' tensor 
$\tilde{T}(k)_{\alpha\beta\mu\nu}$ in the projector basis given by~(\ref{eq:bc3}) and~(\ref{eq:bc4}):
\begin{equation}
\tilde{T}(k)_{\alpha\beta\mu\nu}=\sum_{j=1}^{6}\, x_{j}\,Q(k)^{\sss(j)}_{\alpha\beta,\mu\nu}\,.
\label{eq:bc7}
\end{equation}
Comparing~(\ref{eq:bc6}) with~(\ref{eq:bc7}), we find the relations between the $x_{j}$ coefficients and the $A,\ldots,F$
coefficients:
\begin{gather}
x_{1}=2(C+E)\,,\quad x_{2}=2E\,,\quad x_{3}=2E+3F\,,\nonumber\\
x_{4}=A+2B+4C+2E+F\,,\quad x_{5}=x_{6}=\sqrt{3}(B+F)\,.
\label{eq:bc8}
\end{gather}
The inverse of $\tilde{T}(k)_{\alpha\beta\mu\nu}$ in Eq.~(\ref{eq:bc7})  satisfies
\begin{equation}
\big(\tilde{T}(k)\big)_{\alpha\beta\mu\nu}\,\big(\tilde{T}(k)^{-1}\big)^{\mu\nu}_{\ \ \rho\sigma}=l_{\alpha\beta\rho\sigma}\,,
\label{eq:bc9}
\end{equation}
being $l_{\alpha\beta\rho\sigma}=\big(\eta_{\alpha\mu}\eta_{\beta\nu}+\eta_{\alpha\nu}\eta_{\beta\mu}\big)/2$ the unity
for rank-4 tensors and it  is given by
\begin{equation}
\begin{split}
\big(\tilde{T}(k)^{-1}\big)_{\alpha\beta\mu\nu}&=\Big( +x_{1}^{-1}Q^{\sss(1)}+x_{2}^{-1}Q^{\sss(2)}+
\frac{x_{4}}{\Delta}Q^{\sss(3)}+\\
&\qquad +\frac{x_{3}}{\Delta}Q^{\sss(4)}-\frac{x_{5}}{\Delta}Q^{\sss(5)}-
\frac{x_{6}}{\Delta}Q^{\sss(6)}\Big)(k)_{\alpha\beta\mu\nu}\\
&=:\sum_{j=1}^{6}\, y_{j}\,Q(k)^{\sss(j)}_{\alpha\beta,\mu\nu}\,,
\label{eq:bc10}
\end{split}
\end{equation}
where $\Delta:=x_{3}\cdot x_{4}-x_{5}\cdot x_{6}$. The proof of Eq.~(\ref{eq:bc10}) simply consists in carrying  out the 
product in~(\ref{eq:bc9}) and using the relations of Eq.~(\ref{eq:bc5}). Returning back to the standard representation 
and multiplying with $k^4$, we obtain
\begin{equation}
\begin{split}
\big({T}(k)^{-1}\big)_{\alpha\beta\mu\nu}=\Big[& -2y_{1}+\frac{2y_{2}}{3}+\frac{y_{3}}{3}+y_{4}
-\frac{2y_{5}}{\sqrt{3}},+\frac{y_{5}}{\sqrt{3}}-\frac{y_{3}}{3}+\frac{y_{2}}{3},\\
&+\frac{y_{1}}{2}-\frac{y_{2}}{2},+\frac{y_{2}}{2},
+\frac{y_{3}}{3}-\frac{y_{2}}{3}\Big](k)_{\alpha\beta\mu\nu}\,.
\label{eq:bc11}
\end{split}
\end{equation}
Using the definitions of the $y_{j}$'s as a functions of the $x_{i}$'s from~(\ref{eq:bc10}) and the inverses of the  
relations in~(\ref{eq:bc8}), we can then find the coefficients of the inverse  
$\big(\tilde{T}(k)^{-1}\big)_{\alpha\beta\mu\nu}$ in terms of the original coefficients $A,\ldots,F$.

\addcontentsline{toc}{subsection}{Appendix 4: Numerical Distributions in Eq.~(\ref{eq:b34})}
\section*{Appendix 4: Numerical Distributions in Eq.~(\ref{eq:b34}) }
\renewcommand{\theequation}{D.\arabic{equation}}
\setcounter{equation}{0}

In this appendix we present explicitly the numerical distributions appearing in Eq.~(\ref{eq:b34}) or in 
Eq.~(\ref{eq:b35}). We separate  the various contributions according to their singular order and to the 
presence of graviton or ghost loops.

For the graviton loop, the distribution  with singular order four reads 
\begin{equation}
\hat{d}_{2}^{\sss (4)}(p)_{\alpha\beta\mu\nu}^{\ \,\, \cdot |\cdot}=\frac{\kappa^2 \pi}{960 (2\pi)^4}\,
\Big[ -80,+60,+10,-70,0\Big]\,\Theta(p^2)\,\mathrm{sgn}(p^0)\,.
\end{equation}
The distribution with singular order two can be written as
\begin{equation}
\hat{d}_{2}^{\sss (2)}(p)_{\alpha\beta\mu\nu}^{\ \,  \gamma |\rho}=\frac{ \kappa^2 \pi}{960 (2\pi)^4}\,
\Big[-10\, \hat{P}_{\sss I}(p)_{\alpha\beta\mu\nu}^{\ \, \gamma|\rho} -10\,p^2\,
\hat{P}_{\sss II}(p)_{\alpha\beta\mu\nu}^{\ \,  \gamma|\rho}\Big]\,\Theta(p^2)\,\mathrm{sgn}(p^0)\,,
\end{equation}
with
\begin{equation}
\begin{split}
\hat{P}_{\sss I}(p)_{\alpha\beta\mu\nu}^{\ \, \gamma|\rho}=
&+\big\{ p_{\alpha} p_{\beta}\,\big(8\,\eta_{\mu}^{\ \rho} \eta_{\nu}^{\ \gamma} +8\,\eta_{\mu}^{\ \gamma} \eta_{\nu}^{\ \rho}\big)
                                   +(\alpha\leftrightarrow\beta,\mu\leftrightarrow\nu)\big\}\\
&+\big\{ p_{\alpha} p_{\mu}\,\big(16\, \eta_{\beta\nu} \eta^{\gamma\rho}+12\, \eta_{\beta}^{\ \gamma}\eta_{\nu}^{\ \rho}
         +10\, \eta_{\beta}^{\ \rho}\eta_{\nu}^{\ \gamma}\big)+(\alpha\leftrightarrow\beta)+(\mu\leftrightarrow\nu)\\
                 &\qquad\qquad +(\alpha\leftrightarrow\beta,\mu\leftrightarrow\nu)\big\}\\
&+p^{\rho} p^{\gamma}\,\big\{-34\, \eta_{\alpha\beta} \eta_{\mu\nu}+60\, \eta_{\alpha\mu} \eta_{\beta\nu} 
                                                                +60\, \eta_{\alpha\nu} \eta_{\beta\mu} \big\}\\
&+\big\{4\,p_{\alpha}p^{\gamma}\,\big(-\eta_{\beta\mu} \eta_{\nu}^{\ \rho}-\eta_{\beta\nu} \eta_{\mu}^{\ \rho}+
                   \eta_{\beta}^{\ \rho}\eta_{\mu\nu}\big)  +(\alpha\leftrightarrow\beta)\\
                   &\qquad\qquad+   (\alpha\leftrightarrow\mu,\gamma\leftrightarrow\rho,\beta\leftrightarrow\nu)+
                      (\alpha\leftrightarrow\nu,\gamma\leftrightarrow\rho,\beta\leftrightarrow\mu)\big\}\\
&+\big\{p_{\alpha} p^{\rho}\,\big(-30\,\eta_{\beta\mu}\eta_{\nu}^{\ \gamma}
                                 -30\,\eta_{\beta\nu}\eta_{\mu}^{\ \gamma}
                                 +18\,\eta_{\beta}^{\ \gamma}\eta_{\mu\nu} \big)+(\alpha\leftrightarrow\beta)\\
                   &\qquad\qquad+   (\alpha\leftrightarrow\mu,\gamma\leftrightarrow\rho,\beta\leftrightarrow\nu)+
                      (\alpha\leftrightarrow\nu,\gamma\leftrightarrow\rho,\beta\leftrightarrow\mu)\big\}\,,
\raisetag{30mm}
\end{split}
\end{equation}
and
\begin{equation}
\begin{split}
\hat{P}_{\sss II}(p)_{\alpha\beta\mu\nu}^{\ \,\gamma|\rho}=
&+  \eta_{\alpha\beta} \eta_{\mu\nu} \eta^{\gamma\rho} -8\, \eta^{\gamma\rho}\big\{\eta_{\alpha\mu} \eta_{\beta\nu} 
                         +(\mu\leftrightarrow\nu)     \big\}\\
&-3\, \big\{\eta_{\alpha\beta} \eta_{\mu}^{\ \gamma} \eta_{\nu}^{\ \rho}      +(\mu\leftrightarrow\nu)     \big\}
-3\, \big\{\eta_{\mu\nu} \eta_{\alpha}^{\ \gamma} \eta_{\beta}^{\ \rho}      +(\alpha\leftrightarrow\beta)\big\}\\
+  \big\{2\,\eta_{\alpha}^{\ \gamma} \eta_{\beta\mu} \eta_{\nu}^{\ \rho}&+5 \,\eta_{\alpha\mu} \eta_{\nu}^{\ \gamma} \eta_{\beta}^{\ \rho}+
         (\mu\leftrightarrow\nu)+(\alpha\leftrightarrow\beta)+(\alpha\leftrightarrow\beta,\mu\leftrightarrow\nu)\big\}\,.
\end{split}
\end{equation}
This distribution yields after distribution splitting and after having multiplied it  with $p_{\gamma}p_{\rho}$ the 
following contribution to the graviton self-energy through graviton loop:
\begin{equation}
p_{\gamma}\,p_{\rho}\,\hat{t}_{2}^{\sss (2)}(p)_{\alpha\beta\mu\nu}^{\ \, \gamma|\rho}=\frac{\kappa^2 \pi}{960 (2\pi)^4}\,
\Big[-1200,-380,+450,-520,+330\Big]\,\hat{t}(p)\,.
\end{equation}
The distribution with singular order three attached to $:\!h^{\alpha\beta}(x)_{,\gamma}h^{\mu\nu}(y)\!:$ reads
\begin{equation}
\hat{d}_{2}^{\sss (3a)}(p)_{\alpha\beta\mu\nu}^{\ \,\gamma|\cdot}=\frac{i \kappa^2 \pi}{960 (2\pi)^4}\,
\Big[\hat{Q}_{\sss I}(p)_{\alpha\beta\mu\nu}^{\ \, \gamma|\cdot} +p^2\,\hat{Q}_{\sss II}(p)_{\alpha\beta\mu\nu}^{\ \, \gamma|\cdot}
\Big]\,
\Theta(p^2)\,\mathrm{sgn}(p^0)\,,
\end{equation}
with
\begin{equation}
\begin{split}
\hat{Q}_{\sss I}(p)_{\alpha\beta\mu\nu}^{\ \, \gamma|\cdot}=&
-120\, p_{\mu} p_{\nu}\, \big\{ p_{\alpha}\,\eta_{\beta}^{\ \gamma}+(\alpha\leftrightarrow\beta)\big\}\\
&+20\, \big\{ \eta_{\alpha\beta}\,   p_{\mu} p_{\nu} p^{\gamma}
          +\eta_{\mu}^{\ \gamma}\,p_{\alpha} p_{\beta} p_{\nu} +\eta_{\nu}^{\ \gamma}\,p_{\alpha} p_{\beta} p_{\mu}    \big\}\,,
\end{split}
\end{equation}
and
\begin{equation}
\begin{split}
\hat{Q}_{\sss II}(p)_{\alpha\beta\mu\nu}^{\ \, \gamma|\cdot}=&+p^{\gamma}\,\big\{
+110\,  \eta_{\mu\nu} \eta_{\alpha\beta}
-210\,  \eta_{\mu\alpha} \eta_{\nu\beta}
-210\,  \eta_{\mu\beta} \eta_{\alpha\nu}\big\}\\
&+\big\{p_{\alpha}\,( -20\, \eta_{\mu\nu}    \eta_{\beta}^{\ \gamma}+130\, \eta_{\mu\beta}  \eta_{\nu}^{\ \gamma} 
                                                   +130\,\eta_{\mu}^{\ \gamma} \eta_{\beta\nu})+(\alpha\leftrightarrow\beta) \big\}\\
&+\big\{p_{\mu}\,( -20\, \eta_{\alpha\beta}  \eta_{\nu}^{\ \gamma} +20\, \eta_{\alpha\nu}  \eta_{\beta}^{\ \gamma}
                            +20\,\eta_{\alpha}^{\ \gamma}  \eta_{\nu\beta})+(\mu\leftrightarrow\nu) \big\}\,.
\end{split}
\end{equation}
This distribution yields after distribution splitting and after having multiplied it  with $i\,p_{\gamma}$ the following 
contribution to the graviton self-energy through graviton loop:
\begin{equation}
i\,p_{\gamma}\,\hat{t}_{2}^{\sss (3a)}(p)_{\alpha\beta\mu\nu}^{\ \, \gamma|\cdot}=\frac{\kappa^2 \pi}{960 (2\pi)^4}\,
\Big[+200,+30,-150,+210,-110 \Big]\,\hat{t}(p)\,.
\end{equation}
The distribution $\hat{d}_{2}^{\sss (3b)}(p)_{\alpha\beta\mu\nu}^{\ \,\, \cdot|\rho}$ with singular order three attached to 
the operator $:\!h^{\alpha\beta}(x)h^{\mu\nu}(y)_{,\rho}\!:$ is the same to the previous one with a global sign change 
and the replacements $\mu\leftrightarrow\alpha$, $\nu\leftrightarrow\beta$ and $\rho\leftrightarrow\gamma$. After 
distribution splitting and multiplication with $-i p_{\rho}$, it gives the same  contribution to the graviton self-energy 
through graviton loop as the previous one.

For the ghost loop we have: with singular order four
\begin{equation}
\hat{d}_{2}^{\sss (4)}(p)_{\alpha\beta\mu\nu}^{\ \,\, \cdot |\cdot}=\frac{\kappa^2 \pi}{960 (2\pi)^4}\,
\Big[ +64,-8,+12,+8,+8\Big]\,\Theta(p^2)\,\mathrm{sgn}(p^0)\,,
\end{equation}
after symmetrization in $(\alpha\beta)\leftrightarrow (\mu\nu)$; with singular order three we have
\begin{equation}
\begin{split}
\hat{d}_{2}^{\sss (3b)}(p)_{\alpha\beta\mu\nu}^{\ \,\,  \cdot |\rho}=\frac{i \kappa^2 \pi}{960 (2\pi)^4}\,\Big[
&+20\,p_{\alpha} p_{\beta}\, \big\{ p_{\mu}\,\eta_{\nu}^{\ \rho}+ (\mu\leftrightarrow\nu)\big\}\\
&+20\, p^2\,\big\{ p_{\alpha}\, \eta_{\mu\nu}\eta_{\beta}^{\ \rho}+(\alpha\leftrightarrow\beta)\big\}\\
+10\, p^2\,\big\{ p_{\mu}\,( \eta_{\alpha\beta}\eta_{\nu}^{\ \rho}&-\eta_{\nu\beta}\eta_{\alpha}^{\ \rho}
                       -\eta_{\alpha\nu}\eta_{\beta}^{\ \rho}) +(\mu\leftrightarrow\nu)\big\}
\Big]\,\Theta(p^2)\,\mathrm{sgn}(p^0)\,,
\raisetag{18mm}
\end{split}
\end{equation}
whereas the distribution $\hat{d}_{2}^{\sss (3a)}(p)_{\alpha\beta\mu\nu}^{\ \, \gamma|\cdot}$ with singular order three 
attached to the operator $:\!h^{\alpha\beta}(x)_{,\gamma}h^{\mu\nu}(y)\!:$ is the same to the previous one with a global 
sign change and the replacements $\mu\leftrightarrow\alpha$, $\nu\leftrightarrow\beta$ and $\rho\leftrightarrow\gamma$. 
With singular order two we obtain
\begin{equation}
\hat{d}_{2}^{\sss (2)}(p)_{\alpha\beta\mu\nu}^{\ \, \gamma|\rho}=\frac{\kappa^2 \pi}{960 (2\pi)^4}\,
\Big[+\eta_{\beta}^{\ \rho}\,\eta_{\nu}^{\ \gamma}\,\big( 80\,p_{\alpha}p_{\mu} +40\,p^2\,\eta_{\alpha\mu}\big)\Big]\,
\Theta(p^2)\,\mathrm{sgn}(p^0)\,,
\end{equation}
which must be symmetrized in $(\alpha\leftrightarrow\beta)$ and $(\mu\leftrightarrow\nu)$ and then one applies  $p_{\gamma}p_{\rho}$.

\end{document}